\documentclass[leqno]{article}

\usepackage[]{amsfonts} \usepackage[]{amsmath}
\usepackage[]{amssymb} \usepackage[]{latexsym}
\usepackage{graphicx,color} \usepackage{amsthm}
 \usepackage{url} \usepackage{mathtools}
\usepackage{amssymb}
\usepackage{mathrsfs} \usepackage{graphicx}
\usepackage{subcaption} \usepackage{appendix}
\usepackage{cancel} \usepackage[sort]{natbib}
\usepackage{datatool}

\numberwithin{equation}{section}
\allowdisplaybreaks 

\usepackage{anysize}
\marginsize{2.5cm}{2.5cm}{1.5cm}{4cm}

\numberwithin{equation}{section}

\theoremstyle{plain}
\newtheorem{theorem}{Theorem}[section] \newtheorem*{theorem*}{Theorem}
 \newtheorem*{proposition*}{Proposition}
 \newtheorem*{lemma*}{Lemma}
 \newtheorem*{corollary*}{Corollary}

\theoremstyle{definition}
 \newtheorem*{definition*}{Definition}
 \newtheorem*{example*}{Example}
\newtheorem{remark}[theorem]{Remark} \newtheorem*{remark*}{Remark}
\newtheorem{generalization}{Generalization}
 \newtheorem{assumption}[theorem]{Assumption}

\newcommand{\ds}{\displaystyle} 
\newcommand{\eps}{\varepsilon}
\newcommand{\bE}{\mathbb{E}}

\newcommand{\cF}{\mathcal{F}}
\newcommand{\cA}{\mathcal{A}}
\newcommand{\cM}{\mathcal{M}}

\newcommand{\cL}{\mathcal{L}}

\newcommand{\bR}{\mathbb{R}}
\newcommand{\VIX}{\mbox{VIX}}
\newcommand{\erf}{\mbox{erf}}

\newcommand{\cLcir}{\mathcal{L}_{\!_{C\!I\!R}}}
\newcommand{\rhor}{\raisebox{1.5pt}{$\rho$}}
\newcommand{\varphir}{\raisebox{1.5pt}{$\varphi$}}

\newcommand{\spx}{S\&P 500}
\DeclareMathOperator*{\argmin}{arg\,min}

\begin{document}

\title{Heston Stochastic Vol-of-Vol Model for Joint Calibration of VIX and \spx\ Options}

\author{Jean-Pierre Fouque\thanks{Department of Statistics \& Applied Probability, University of California, Santa Barbara, CA 93106-3110,
{\em fouque@pstat.ucsb.edu}. Work supported by NSF grant DMS-1107468.} ,
Yuri F. Saporito\thanks{Escola de Matem\'atica Aplicada (EMAp), Funda\c{c}\~ao Get\'ulio Vargas (FGV), Rio de Janeiro, RJ 22250-900, Brazil,
{\em yuri.saporito@fgv.br}.}}

\maketitle

\abstract{A parsimonious generalization of the Heston model is proposed where the volatility-of-volatility is assumed to be stochastic. We follow the perturbation technique of Fouque \textit{et al} (2011, CUP) to derive a first order approximation of the price of options on a stock and its volatility index. This approximation is given by Heston's quasi-closed formula and some of its Greeks. It can be very efficiently calculated since it requires to compute only Fourier integrals and the solution of simple ODE systems. We exemplify the calibration of the model with \spx\ and VIX data.}


\section{Introduction}

The volatility index of the \spx, acronymed VIX and also known as the \textit{fear index}, has drawn the attention of researchers and practitioners alike since its first introduction in the US market in 1993, see \cite{cboe03} for its current definition. In 2004, future contracts on VIX began to trade at CBOE Futures Exchange and later on, in 2006, options on VIX were firstly negotiated.

From its definition, the VIX index is computed using the price of liquid options on the \spx. In fact, 
\begin{align}
\VIX_t^2 \approx \frac{2}{\tau_0} \int_0^{+\infty} Q(t,t+\tau_0, K) \frac{dK}{K^2}, \label{eq:vix_def}
\end{align}
where $\tau_0$ is 30 days and $Q(t,T,K)$ denotes the price of the out-the-money option at time $t$ with maturity $T$ and strike $K$. The approximation sign $\approx$ appears in the equation above because, obviously, the index is computed by discretizing the integral on the left-hand side. Moreover, the square of the VIX is linearly interpolated between the two closest maturities in order to have its value 30 days from $t$.

Hence, the implied volatility surfaces of the \spx\ and VIX are highly connected, and as a consequence this dependence is very complex. There are few models proposed to solve this calibration issue, see, for instance, \cite{joint_3_2_jump, joint_carr_madan_jump, joint_heston_plus_plus, joint_regime_switching} and \cite{joint_cont_calibration}. The common aspect of the models described in these references is the presence of jumps in the stock price and/or its spot volatility. Differently from these models, we consider here a continuous diffusion model. More precisely, we propose a simple generalization of \cite{heston93} where volatility-of-volatility is stochastic. 

Furthermore, to the best of our knowdelege, the only continuous models proposed to joint calibrate stock and volatility options appeared in \cite{joint_gatheral}. While the calibration of these models rely on Monte Carlo or PDE methods, ours grants us quasi-closed formulas for the first-order approximation of option prices on both markets. Additionally, in the direction of model-free results, there is the work \cite{joint_de_marco_labordere}.

Our approach is based on the multiscale stochastic volatility perturbation technique proposed by Fouque, Papanicolaou, Sircar and S\o lna, see \cite{multiscale_fouque_new_book}. This approach allows us to approximate option prices under the full model by their prices and Greeks under a simpler model. This method is very flexible and can be adapted to a large number of models and options, as one will be able to see in this paper. 

A different perturbation technique was applied in the context of joint calibration in \cite{joint_regime_switching}. In this paper, the authors proposed a regime-switching generalization of the Heston model. The perturbation is done in the jump-process that brings the regime-switching feature into the model. This is fundamentally different from the solution proposed here.

Additionally, the Heston model was also generalized in the lines of the multiscale stochastic volatility modelling in \cite{fouque_fast_heston}. This generalization is also very different from the one pursued here. The main issue is that the simple formula found in our model, shown in Equation (\ref{eq:vix_formula}), is not verified in the aforesaid model.

The paper is organized as follows: we describe our model in Section \ref{sec:model} and the main results are stated in Section \ref{sec:results}. We discuss the calibration of the proposed model and exemplify it in Section \ref{sec:calibration_all}. Some generalizations of our model are outlined in Section \ref{sec:generalization}. Finally, the rationale and computations that justify our first-order approximation are shown in Section \ref{sec:perturbation} and in the Appendices \ref{sec:fourier_S} and \ref{sec:fourier_vix}. 

\section{The Model}\label{sec:model}

We will assume that the stock price $S$, under a risk-neutral probability, follows a Heston dynamics with a stochastic volatility-of-volatility (vol-vol):
\begin{align}\label{eq:sde_risk_neutral}
\left\{
\begin{array}{l}
  dS_t = (r - q)S_t dt + \sqrt{V_t} S_t dW_t^S, \\ \\
  dV_t = \kappa(m - V_t)dt + \eta_t \sqrt{V_t} dW_t^V, \\ \\
  dW_t^SdW_t^V = \rhor_{SV} dt
\end{array}
\right.
\end{align}
where $\eta_t$ will be specified later in this section. We will denominate this model by Heston SVV model.

The volatility index of $S$ at $t$, which it will be denoted, because of obvious reasons, by $\VIX_t$, is defined as
\begin{align}
\VIX_t^2 = \bE\left[\left.\frac{1}{\tau_0} \int_t^{t+\tau_0} V_u du \ \right| \ \cF_t \right], \label{eq:VIX}
\end{align}
where $\tau_0 = 30/360$, i.e. 30 calendar days. The main example to have in mind is the S\&P 500 and the VIX. The expected value above, as all the other expected values in this work, is under the chosen risk-neutral measure. This risk-neutral measure is taken to match the vanilla option prices for both the stock and its volatility index markets.

\begin{remark}

As it was discussed in the introduction, the VIX is computed by discretizing Equation (\ref{eq:vix_def}). However, under any continuous model with spot variance $V$, the left-hand side of Equation (\ref{eq:vix_def}) can be written as Equation (\ref{eq:VIX}). Hence, we are actually incurring in a small discretization error when using the non-discretized version of the volatility index.

\end{remark}

Let us now specify the particular formula for the process $\eta_t$. In order to be able to find a computationally efficient approximation for the price of options on $S$ and VIX, we choose $\eta_t$ to be governed by fast and slow time scales. More precisely,
\begin{align}\label{eq:sde_risk_neutral_multiscale}
\left\{
\begin{array}{l}
  \eta_t = \eta(Y^{\eps}_t,Z^{\delta}_t) \\ \\
  \ds dY^{\eps}_t = \frac{V_t}{\eps} \alpha(Y_t^{\eps})dt + \sqrt{\frac{V_t}{\eps}} \beta(Y_t^{\eps}) dW_t^Y, \\ \\
  dZ^{\delta}_t = V_t \delta c(Z^{\delta}_t)dt + \sqrt{\delta V_t} g(Z^{\delta}_t) dW_t^Z,
\end{array}
\right.
\end{align}
where $(W_t^S,W_t^V,W_t^Y,W_t^Z)$ is a correlated Brownian motion with
$$dW_t^idW_t^j = \rhor_{ij} dt, \ i,j=S,V,Y,Z.$$

\begin{assumption}\label{ass:model}

The assumptions of this model are:
\begin{itemize}

\item there exists a unique strong solution of the stochastic differential equations (\ref{eq:sde_risk_neutral})-(\ref{eq:sde_risk_neutral_multiscale}) for fixed $(\eps,\delta)$;

\item the covariance matrix of $(W_t^S,W_t^V,W_t^Y,W_t^Z)$ is positive-definite;

\item $\alpha$ and $\beta$ are such that the process $Y^1$ has a unique invariant distribution and is mean-reverting as in \cite[Section 3.2]{multiscale_fouque_new_book};

\item $\eta(y,z)$ is a positive function, smooth in $z$ and such that $\eta^2(\cdot,z)$ is integrable with respect to the invariant distribution of $Y^1$.

\end{itemize}

\end{assumption}

\begin{remark}[Feller Condition]

In order to guarantee that $V_t > 0$ a.s. for all $t \in [0,T]$, one needs to assume
$$2\kappa m \geq \eta(y, z) \geq c > 0,$$
for every $(y,z)$ and a constant $c > 0$. See, for example, \cite{time_heston}.

\end{remark}

\section{Main Results}\label{sec:results}

In this section we present the first-order approximation for the price of derivative contracts on the stock price, $S$, and on its volatility index, VIX. The derivation of these results and formulas will be fully developed in the sections to follow. 

We start by fixing two European derivatives with maturity $T$ and payoff functions $\varphir_S$ and $\varphir_V$ that depend only on the terminal values $S_T$ and $\VIX_T$, respectively. The no-arbitrage prices, under the chosen risk-neutral measure, of these derivative contracts are given by the conditional expectations:
\begin{align}
P^{\eps,\delta}_S(t,s,v,y,z) &= \bE[e^{-r(T-t)}\varphir_S(S_T) \ | \ S_t = s, V_t = v, Y^{\eps}_t = y, Z^{\delta}_t = z],\label{eq:opt_price_S}\\
P^{\eps,\delta}_V(t,v,y,z) &= \bE[e^{-r(T-t)}\varphir_V(\VIX_T) \ | \ V_t  = v, Y^{\eps}_t = y, Z^{\delta}_t = z].\label{eq:opt_price_VIX}
\end{align}
In the case of VIX futures (i.e. $\varphir_V(v) = v$), there is no discounting term. 

We are interested in jointly calibrating our model to options on $S$ and on VIX. Below, we present the first-order approximation for these option prices.

\begin{remark}
More precisely, we say that a function $g^{\eps,\delta}$ is a first-order approximation to the function $f^{\eps,\delta}$ if
$$|g^{\eps,\delta} - f^{\eps,\delta}| \leq C(\eps + \delta),$$
for some constant $C > 0$ and for sufficiently small $\eps,\delta > 0$. We use the notation
\begin{align}
g^{\eps,\delta} - f^{\eps,\delta} = O(\eps + \delta). \label{eq:bigO}
\end{align}
\end{remark}

We have then the following theorem, see Section \ref{sec:accuracy}.

\begin{theorem*}[Accuracy Theorem]

Under Assumption \ref{ass:model} and if the payoff functions $\varphir_S$ and $\varphir_V$ are continuous and piecewise smooth, then
\begin{align*}
P^{\eps,\delta}_S(t,s,v,y,z) &= P_{S_0}(t,s,v,z) + P^{\eps}_{S_{1,0}}(t,s,v,z) + P^{\delta}_{S_{0,1}}(t,s,v,z) + O(\eps + \delta),\\
P^{\eps,\delta}_V(t,v,y,z) &= P_{V_0}(t,v,z) + P^{\eps}_{V_{1,0}}(t,v,z) + P^{\delta}_{V_{0,1}}(t,v,z) + O(\eps + \delta).
\end{align*}

\end{theorem*}

More importantly, each of the functions on the right-hand side of the equations above can be efficiently computed. Indeed, for options on $S$, we find that $P_{S_0}$ is the no-arbitrage price of $\varphir_S$ under the Heston model with constant vol-vol equals $\overline{\eta}(z)$ and effective correlation equals $\overline{\rho}(z)$. Specifically, we have
\begin{align*}
&P_{S_0} + P^{\eps}_{S_{1,0}} + P^{\delta}_{S_{0,1}} = \frac{e^{-r \tau}}{\pi} \int_0^{+\infty} Re\left( e^{-i\xi x(t,s)}(1 + h_{S_0}^{\eps,\delta} + v h_{S_1}^{\eps,\delta} + v^2 h_{S_2}^{\eps,\delta}) G_S(\tau,\xi,v,z)\widehat{\varphir_S}(\xi) \right) d\xi_r,
\end{align*}
where
\begin{align*}
&\tau(t) = T - t, \\
&x(t,s) = (r-q)(T-t) + \log s, \\
&\widehat{\varphir_S}(\xi) = \int_{\bR} \varphir_S(e^{x - (r-q)\tau}) e^{-i \xi x} dx, \quad \xi = \xi_r + i\xi_i,\\
&G_S(\tau,\xi,v,z) = e^{C(\tau,\xi,z) + vD(\tau,\xi,z)}, \\
&C(\tau,\xi,z) = \frac{\kappa m }{\overline{\eta}^2(z)} \left( (\kappa + i\overline{\rho}(z) \overline{\eta}(z)\xi - d(\xi, z)) \tau - 2 \log\left( \frac{e^{-d(\xi, z)\tau}/g(\xi, z) - 1}{1/g(\xi, z) - 1} \right) \right), \\
&D(\tau,\xi,z) = \frac{\kappa + i\overline{\rho}(z) \overline{\eta}(z)\xi + d(\xi, z)}{\overline{\eta}^2(z)}\left( \frac{1 - e^{d(\xi, z)\tau}}{1 - g(\xi, z)e^{d(\xi, z)\tau}} \right), \\
&g(\xi, z) = \frac{\kappa + i\overline{\rho}(z) \overline{\eta}(z)\xi + d(\xi, z)}{\kappa + i\overline{\rho}(z) \overline{\eta}(z)\xi - d(\xi, z)},\\
&d(\xi, z) = \sqrt{\overline{\eta}^2(z)(\xi^2 - i\xi) + (\kappa + i\overline{\rho}(z) \overline{\eta}(z)\xi)^2},\\
&h_{S_0}^{\eps,\delta} = f^{\eps}_0 + g^{\delta}_0, \quad h_{S_1}^{\eps,\delta} = f^{\eps}_1 + g^{\delta}_1, \quad h_{S_2}^{\eps,\delta} = g^{\delta}_2,
\end{align*}
with $f^{\eps}_0$ and $f^{\eps}_1$ satisfying the ODE system:
\begin{align*}
\left\{
\begin{array}{l}
\ds \frac{\partial f^{\eps}_1}{\partial \tau}(\tau,\xi,z) =  (\overline{\eta}^2(z)D(\tau,\xi,z) - (\kappa +  \overline{\rho}(z)\overline{\eta}(z) i\xi))f^{\eps}_1(\tau,\xi,z) \\ \\
 \ds \hskip 1cm -i\xi V_{1,2}^{\eps}(z)D^2(\tau,\xi,z) -\xi^2 V_{2,1}^{\eps}(z) D(\tau,\xi,z) +  V_{0,3}^{\eps}(z)D^3(\tau,\xi,z), \\ \\
\ds \frac{\partial f^{\eps}_0}{\partial \tau}(\tau,\xi,z) = \kappa m f^{\eps}_1(\tau,\xi,z),  \\ \\
f^{\eps}_0(0,\xi,z) = f^{\eps}_1(0,\xi,z) = 0,
\end{array}
\right.
\end{align*}
and $g^{\delta}_0$, $g^{\delta}_1$ and $g^{\delta}_2$ satisfying the ODE system:
\begin{align*}
\left\{\begin{array}{l}
\ds \frac{\partial g^{\delta}_2}{\partial \tau}(\tau,\xi,z) = -2(\kappa +  \overline{\rho}(z)\overline{\eta}(z) i\xi - \overline{\eta}^2(z) D(\tau,\xi,z)) g^{\delta}_2(\tau,\xi,z) \\ \\
\hskip 1cm \ds (V_{0,1,\eta}^{\delta}(z) -i\xi V_{1,0,\eta}^{\delta}(z) )\frac{\partial D}{\partial \eta} + (V_{0,1,\rho}^{\delta}(z)\ -i\xi V_{1,0,\rho}^{\delta}(z) )\frac{\partial D}{\partial \rho}, \\ \\
\ds \frac{\partial g^{\delta}_1}{\partial \tau}(\tau,\xi,z) = -(\kappa +  \overline{\rho}(z)\overline{\eta}(z) i\xi - \overline{\eta}^2(z) D(\tau,\xi,z)) g^{\delta}_1(\tau,\xi,z) \\ \\
 \ds \hskip 1cm + \overline{\eta}^2(z) g^{\delta}_2(\tau,\xi,z) + (V_{0,1,\eta}^{\delta}(z)-i\xi V_{1,0,\eta}^{\delta}(z)) \frac{\partial C}{\partial \eta} + (V_{0,1,\rho}^{\delta}(z) -i\xi V_{1,0,\rho}^{\delta}(z)) \frac{\partial \overline{C}}{\partial \rho},  \\ \\
\ds \hskip 1cm + V_{0,1,\eta}^{\delta}(z) \frac{\partial D}{\partial \eta} + V_{0,1,\rho}^{\delta}(z) \frac{\partial D}{\partial \rho} \\ \\
\ds \frac{\partial g^{\delta}_0}{\partial \tau}(\tau,\xi,z) = \kappa m  g^{\delta}_1(\tau,\xi,z), \\ \\
 g^{\delta}_0(0,\xi,z) =  g^{\delta}_1(0,\xi,z) =  g^{\delta}_2(0,\xi,z) = 0.
\end{array}
\right.
\end{align*}
The market group parameters $(\overline{\eta}(z), \overline{\rho}(z), V_{1,2}^{\eps}(z), V_{2,1}^{\eps}(z), V_{0,3}^{\eps}(z), V_{1,0,\eta}^{\delta}(z),V_{0,1,\eta}^{\delta}(z)$, $V_{1,0,\rho}^{\delta}(z), V_{0,1,\rho}^{\delta}(z))$ are related to the functions describing the model (\ref{eq:sde_risk_neutral})-(\ref{eq:sde_risk_neutral_multiscale}) through the equations:
\begin{align}
&\overline{\eta}(z) = \sqrt{\langle \eta^2(\cdot,z) \rangle}, \label{eq:eta_bar_intro} \\
&\overline{\rho}(z) = \rhor_{SV} \frac{\langle \eta(\cdot,z) \rangle}{\overline{\eta}(z)}, \label{eq:rho_bar_intro}\\
&V_{1,2}^{\eps}(z) = -\sqrt{\eps}\frac{ \rhor_{SY} }{2} \left\langle \beta\frac{\partial \phi}{\partial y}(\cdot,z)  \right\rangle - \sqrt{\eps}\rhor_{SV} \rhor_{VY} \left\langle \eta(\cdot, z) \beta \frac{\partial \psi}{\partial y}(\cdot,z) \right\rangle, \label{eq:V12_eps_intro}\\
&V_{2,1}^{\eps}(z) = - \sqrt{\eps}\rhor_{SV} \rhor_{SY} \left\langle \beta\frac{\partial \psi}{\partial y}(\cdot,z)  \right\rangle, \label{eq:V21_eps_intro}\\
&V_{0,3}^{\eps}(z) = -\sqrt{\eps}\frac{ \rhor_{VY} }{2} \left\langle \eta(\cdot, z) \beta\frac{\partial \phi}{\partial y}(\cdot,z)  \right\rangle,\label{eq:V03_eps_intro} \\
&V_{0,1,\eta}^{\delta}(z) = \sqrt{\delta} \rhor_{VZ} g(z) \langle \eta(\cdot,z) \rangle \overline{\eta}'(z), \label{eq:V01eta_delta_intro}\\
&V_{0,1,\rho}^{\delta}(z) = \sqrt{\delta} \rhor_{VZ} g(z) \langle \eta(\cdot,z) \rangle \overline{\rho}'(z), \label{eq:V01rho_delta_intro}\\
&V_{1,0,\eta}^{\delta}(z) = \sqrt{\delta} \rhor_{SZ} g(z)\overline{\eta}'(z), \label{eq:V10eta_delta_intro}\\
&V_{1,0,\rho}^{\delta}(z) = \sqrt{\delta} \rhor_{SZ} g(z)\overline{\rho}'(z).\label{eq:V10rho_delta_intro}
\end{align}

Now, for options on VIX, we find
\begin{align*}
P_{V_0} + P_{V_{1,0}}^{\eps} + P_{V_{0,1}}^{\delta} = \frac{e^{-r\tau}}{\pi} \int_0^{+\infty} Re\left((1 + h_{V_0}^{\eps,\delta} + v h_{V_1}^{\eps,\delta} + v^2 h_{V_2}^{\eps,\delta}) G_V(t,v,\nu)  \widehat{\varphir_V}(\nu) \right) d\nu_i,
\end{align*}
where
\begin{align*}
&\widehat{\varphir_V}(\nu) = \int_0^{+\infty} e^{i \nu v} \varphir_V(\gamma(v))dv, \quad \nu = \nu_r + i \nu_i,\\
&\gamma(v) = \sqrt{m(1 - \theta) + \theta v},\\
&\theta = \frac{1 - e^{-\kappa \tau_0}}{\kappa \tau_0}, \quad \tau_0 = 30/360,\\
&G_V(t,v,\nu,z) = e^{A(\tau,\nu,z) + vB(\tau,\nu,z)}, \\
&A(\tau,\nu,z) = -\frac{2\kappa m}{\overline{\eta}^2(z)} \log\left(\nu\frac{\overline{\eta}^2(z)}{2\kappa}(1-e^{-\kappa \tau}) + 1 \right),\\
&B(\tau,\nu,z) = \frac{\nu e^{-\kappa \tau}}{\nu\frac{\overline{\eta}^2(z)}{2\kappa}(1- e^{-\kappa \tau}) + 1}.
\end{align*}
with $h_{V_0}^{\eps,\delta}$, $h_{V_1}^{\eps,\delta}$ and $h_{V_2}^{\eps,\delta}$ satisfying the ODE system:
\begin{align*}
\left\{\begin{array}{l}
\ds \frac{\partial h_{V_2}^{\eps,\delta}}{\partial \tau}(\tau,\nu,z) = 2(-\kappa + B(\tau,\nu) \overline{\eta}^2(z))h_{V_2}^{\eps,\delta}(\tau,\nu,z) + V_1^{\delta}(z) B(\tau,\nu) \frac{\partial B}{\partial \eta}(\tau,\nu),  \\ \\
\ds \frac{\partial h_{V_1}^{\eps,\delta}}{\partial \tau}(\tau,\nu,z) = (-\kappa + B(\tau,\nu) \overline{\eta}^2(z))h_{V_1}^{\eps,\delta}(\tau,\nu,z) + (2\kappa m + \overline{\eta}^2(z))h_{V_2}^{\eps,\delta}(\tau,\nu,z)\\ \\
 \ds \hskip 1cm + V_3^{\eps}(z) B^3(\tau,\nu) + V_1^{\delta}(z) \left(\frac{\partial B}{\partial \eta}(\tau,\nu) + B(\tau,\nu)\frac{\partial A}{\partial \eta}(\tau,\nu)\right), \\ \\
\ds \frac{\partial h_{V_0}^{\eps,\delta}}{\partial \tau}(\tau,\nu,z) = \kappa m h_{V_1}^{\eps,\delta}(\tau,\nu,z), \\ \\
h_{V_0}^{\eps,\delta}(0,\nu,z) = h_{V_1}^{\eps,\delta}(0,\nu,z) = h_{V_2}^{\eps,\delta}(0,\nu,z) = 0.
\end{array}
\right.
\end{align*}
Moreover, the group market parameters $(V_1^{\delta}(z), V_3^{\eps}(z))$ are given by
\begin{align}
V_3^{\eps}(z) &= -\sqrt{\eps}\frac{ \rhor_{VY} }{2} \left\langle \frac{\partial \phi}{\partial y}(\cdot,z) \eta(\cdot,z)\beta \right\rangle, \label{eq:V3_eps_vix_intro}\\
V_1^{\delta}(z) &= \sqrt{\delta} \rhor_{VZ} g(z)  \langle \eta(\cdot,z) \rangle \overline{\eta}'(z). \label{eq:V1delta_vix_intro}
\end{align}

\section{Calibration}\label{sec:calibration_all}

Firstly, we would like to point out that all the parameters related to the first-order approximation of $P_V^{\eps,\delta}$, i.e. the market group parameters $(\kappa, m, \overline{\eta}(z), V_1^{\delta}(z), V_3^{\eps}(z))$, appear in the first-order approximation of $P_S^{\eps,\delta}$. Indeed, notice that $ V_{0,3}^{\eps}(z) = V_3^{\eps}(z)$ and $V_{0,1,\eta}^{\delta}(z) = V_1^{\delta}(z)$, see Equations (\ref{eq:V03_eps_intro}), (\ref{eq:V3_eps_vix_intro}), (\ref{eq:V01eta_delta_intro}) and (\ref{eq:V1delta_vix_intro}), respectively. The market group parameters $(\overline{\rho}(z), V_{1,2}^{\eps}(z), V_{2,1}^{\eps}(z), V_{1,0,\eta}^{\delta}(z),$ $V_{1,0,\rho}^{\delta}(z), V_{0,1,\rho}^{\delta}(z))$ appear exclusively in the first-order approximation of $P_S^{\eps,\delta}$. \\

Assume there are available $M_S$ and $M_V$ options on $S$ and on VIX, respectively. If we denote all parameters by simply $\Theta$ and the implied volatility of options on $S$ by $\widehat{\sigma}_S^i$ and of options on VIX by $\widehat{\sigma}_V^i$, we will consider the following calibration problem:
\begin{align}\label{eq:calibration}
\widehat{\Theta} = \argmin_{\Theta} \frac{1}{M_S + M_V} \left(M_S\sum_{i=1}^{M_S} (\sigma_S^i(\Theta) - \widehat{\sigma}_S^i)^2 +  M_V \sum_{i=1}^{M_V} (\sigma_V^i(\Theta) - \widehat{\sigma}_V^i)^2\right).
\end{align}

The choice of the initial guess for the optimization problem above is important in order to avoid its many local minima. In this paper, we first consider the standard Heston model and calibrate it to the implied volatility seen in the market, by solving the optimization problem (\ref{eq:calibration}). Then, we use these values and set the $V^{\eps, \delta}$ parameters to zero as the initial guess.

\begin{figure}[h!]
    \centering
    \includegraphics[width=0.7\linewidth]{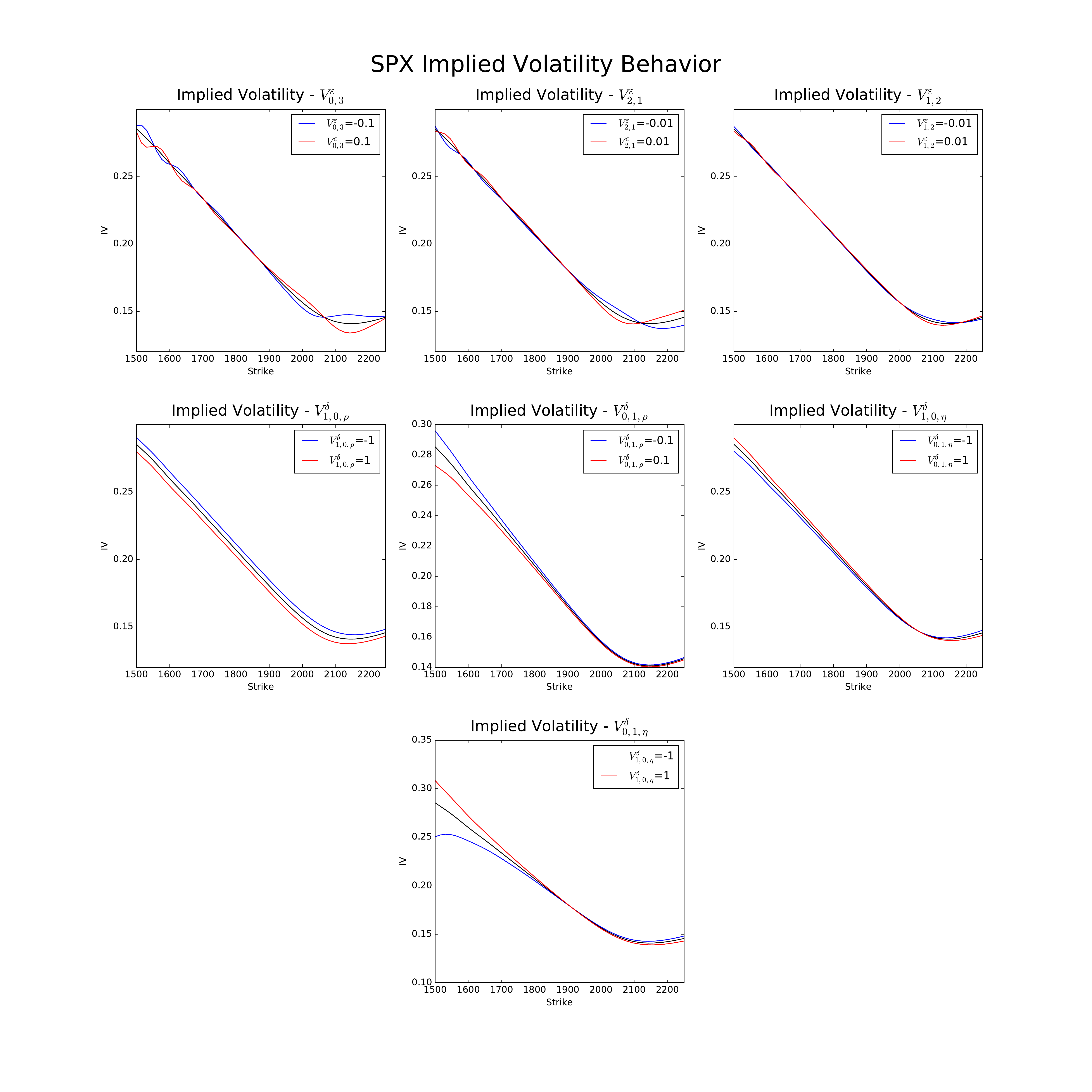}
    \vspace{-1cm}
    \caption{Impact of $V^{\eps,\delta}$'s on \spx's implied volatilities.}
    \label{fig:example_iv_spx}
  \end{figure}

\begin{figure}[h!]
    \centering
    \includegraphics[width=0.6\linewidth]{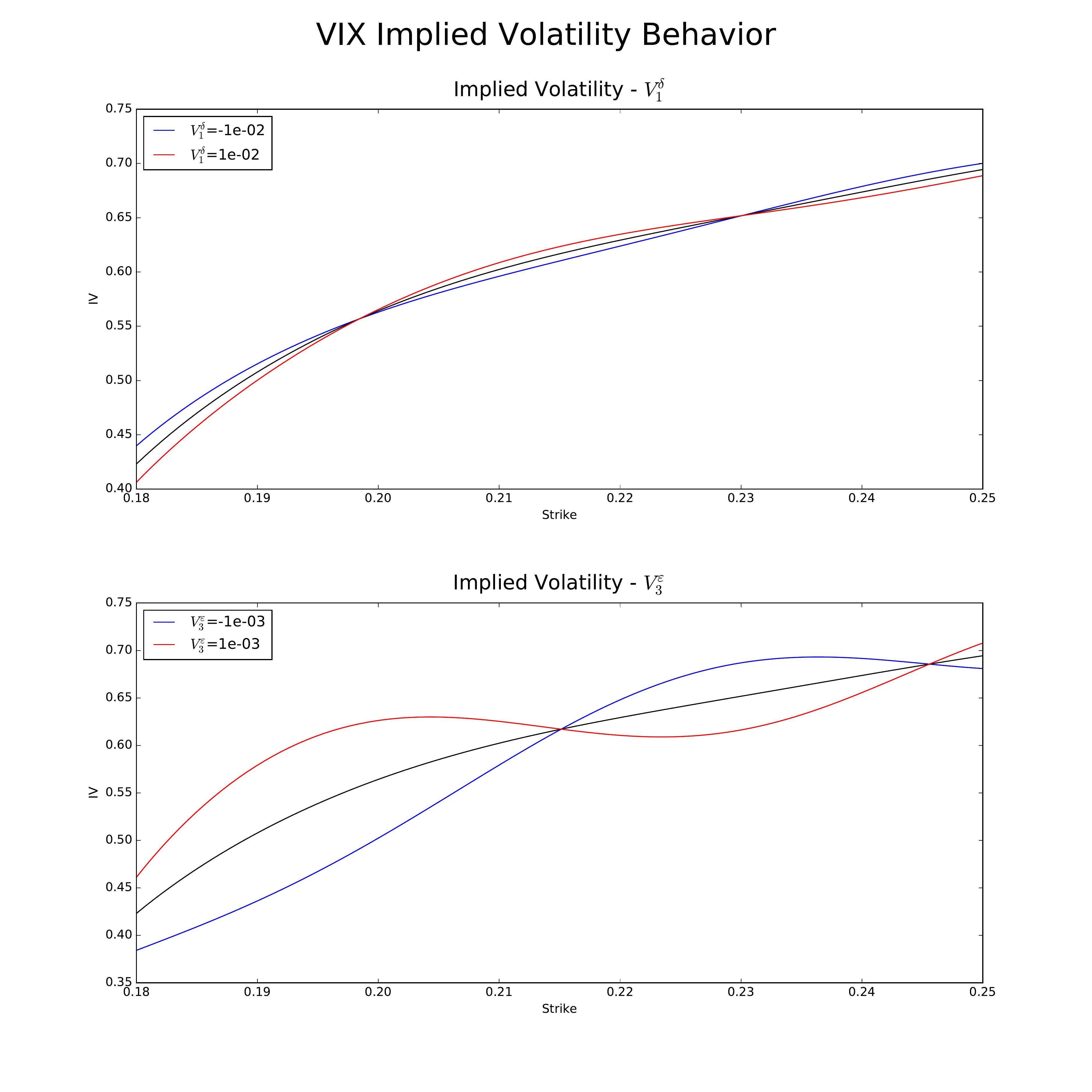}
    \vspace{-1cm}
    \caption{Impact of $V^{\eps,\delta}$'s on VIX's implied volatilities.}
    \label{fig:example_iv_vix}
\end{figure}

We illustrate in Figures \ref{fig:example_iv_spx} and \ref{fig:example_iv_vix} the effect of the $V^{\eps,\delta}$'s on implied volatilities of the SPX and VIX. We used the parameter values $\kappa = 15$, $m=0.04$, $\overline{\eta}(z) = 2.0$, $\overline{\rho}(z) = -0.5$, $S_0 = 2000$ and $\VIX_0 = 0.15$. Interest and dividend rates were set to zero. Moreover, we considered unusually high values of the $V^{\eps,\delta}$'s to accentuate the impact in the implied volatility.

\subsection{Real Data Example}\label{sec:numerical_example}

\begin{figure}[h!]
  \begin{subfigure}{0.5\linewidth}
    \centering
    \includegraphics[width=\linewidth]{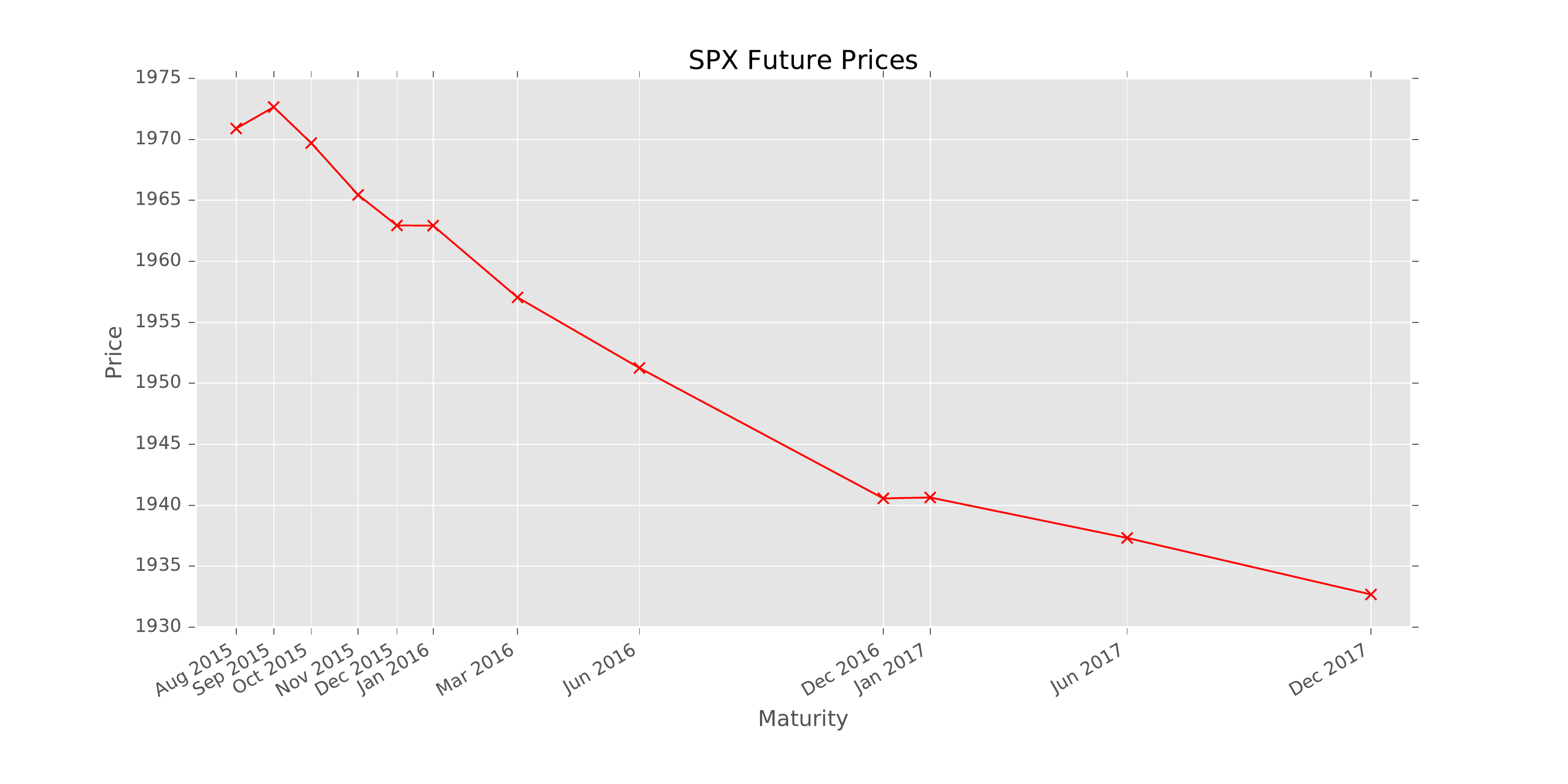}
    \vspace{-0.8cm}
    \caption{\spx}
    \label{fig:future_spx}
  \end{subfigure}
  \hspace{0.5cm}
  \begin{subfigure}{0.5\linewidth}
    \centering
    \includegraphics[width=\linewidth]{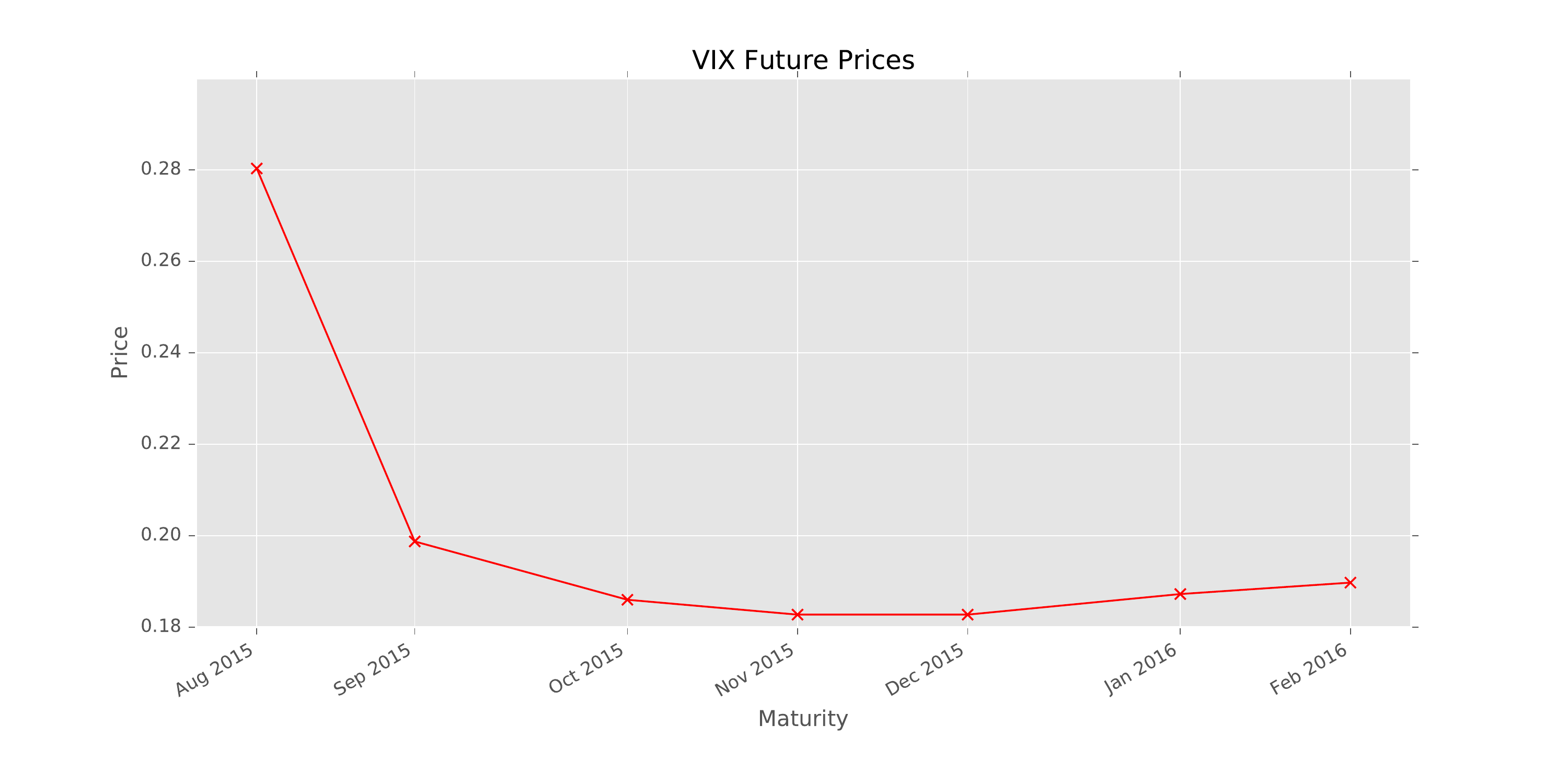}
    \vspace{-0.8cm}
    \caption{VIX}
    \label{fig:future_vix}
  \end{subfigure}
      \caption{Futures Term Structure}
      \label{fig:future}
        \begin{subfigure}{0.5\linewidth}
    \centering
    \includegraphics[width=\linewidth]{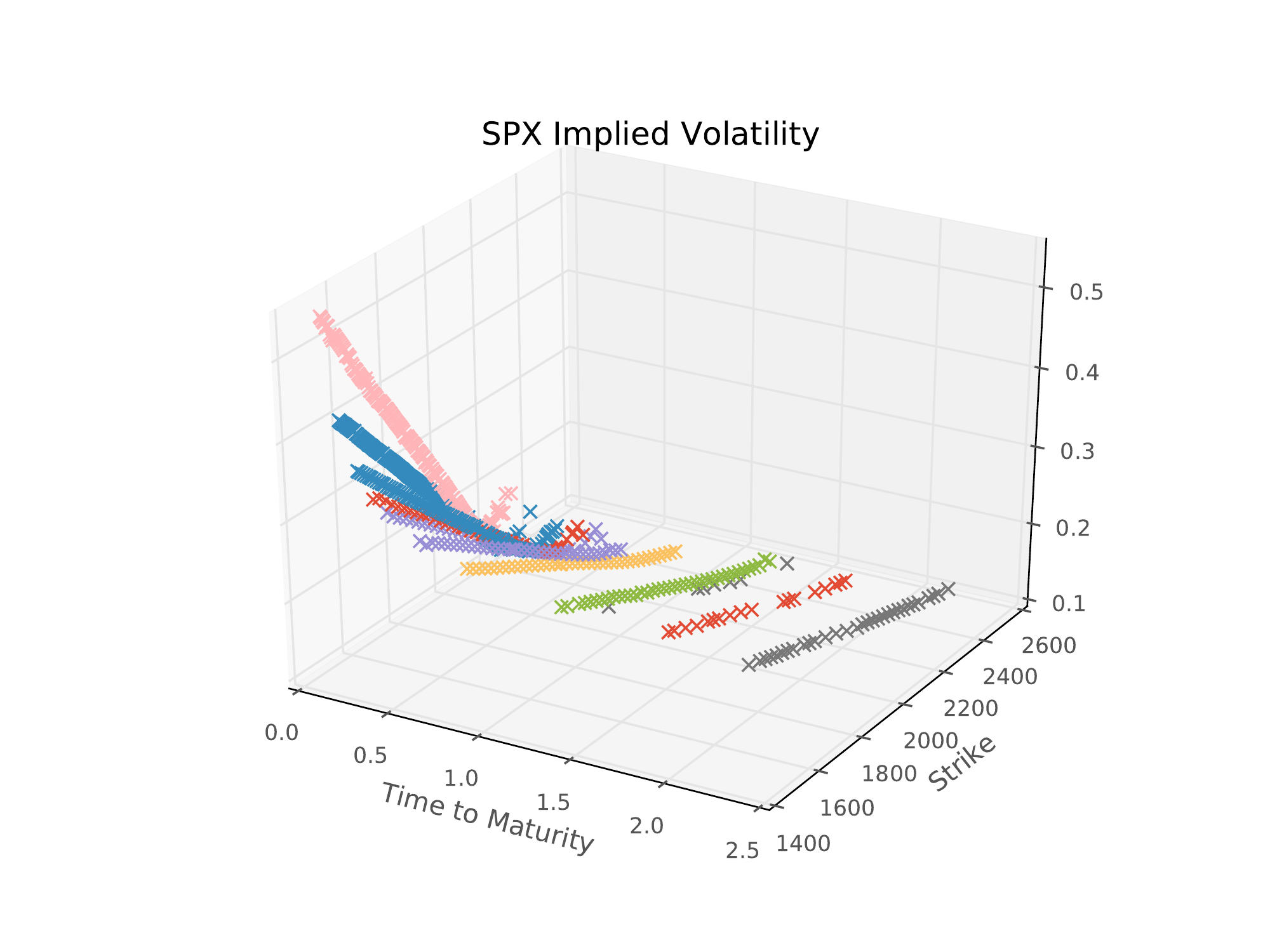}
    \vspace{-0.8cm}
    \caption{\spx}
    \label{fig:iv_spx}
  \end{subfigure}
  \hspace{0.5cm}
  \begin{subfigure}{0.5\linewidth}
    \centering
    \includegraphics[width=\linewidth]{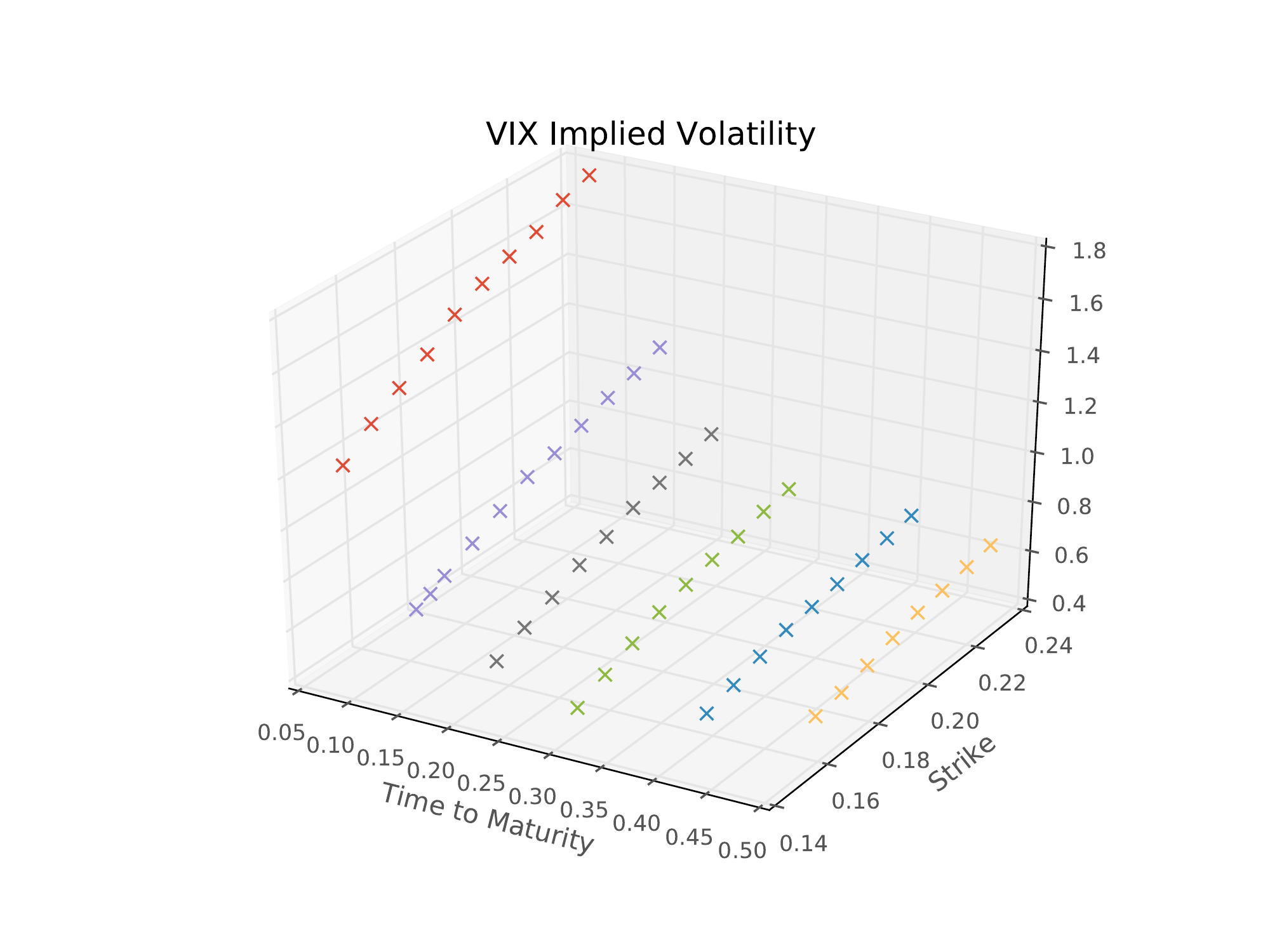}
    \vspace{-0.8cm}
    \caption{VIX}
    \label{fig:iv_vix}
  \end{subfigure}
      \caption{Implied Volatility Surfaces.}
      \label{fig:iv}
      \end{figure}

In this section, we follow the calibration procedure outlined above on real data. We consider implied volatility surfaces of the \spx\ and the VIX on August, 21 of 2015. On this day, the \spx\ closed at 1970.89, and the VIX at 28.03. In order to compute the corresponding future price of \spx\ and VIX, we use the Put-Call Parity with ATM options, see \cite{spx_vix_paper_data}. Implied volatilities are computed using these future prices. See Figures \ref{fig:future} and \ref{fig:iv}.

We cleaned the data based on the adjustments described in \cite{spx_vix_paper_data}. Namely, we removed implied volatilities:

\begin{enumerate}

\item of in-the-money options;

\item with moneyness below 75\% and above 125\%;

\item with zero traded volume;

\item with zero open interest.

\item with maturity larger than 1 year.

\end{enumerate}

In order to understand the improvement of the correction terms of our first-order approximation, we follow the same calibration procedure outlined above for the Heston model (i.e. when the vol-vol is constant).

\begin{figure}[h!]
  \begin{subfigure}{0.5\linewidth}
    \centering
    \includegraphics[width=\linewidth]{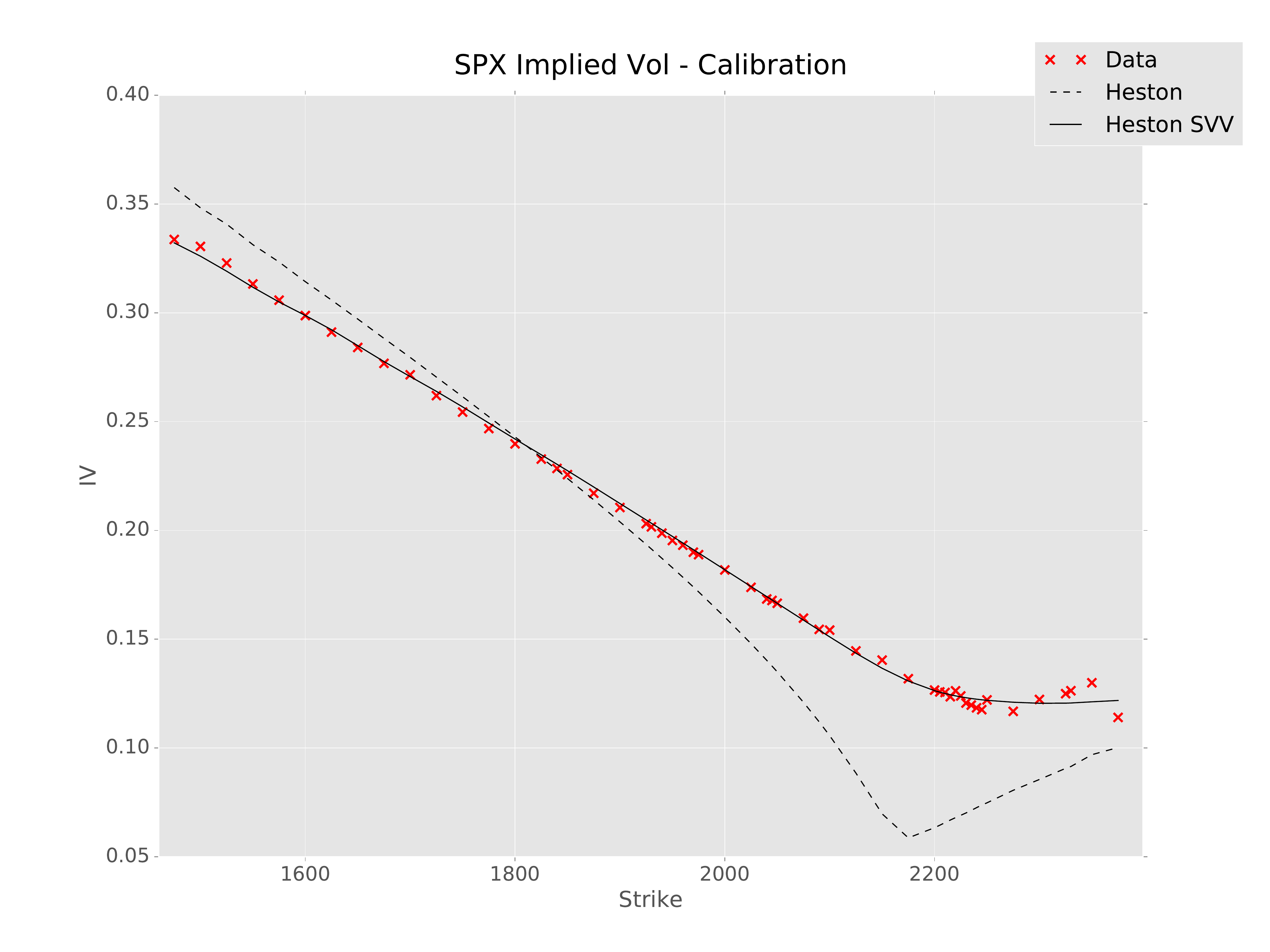}
    \vspace{-0.8cm}
    \caption{\spx}
    \label{fig:calibrated_iv_spx}
  \end{subfigure}
  \hspace{0.5cm}
  \begin{subfigure}{0.5\linewidth}
    \centering
    \includegraphics[width=\linewidth]{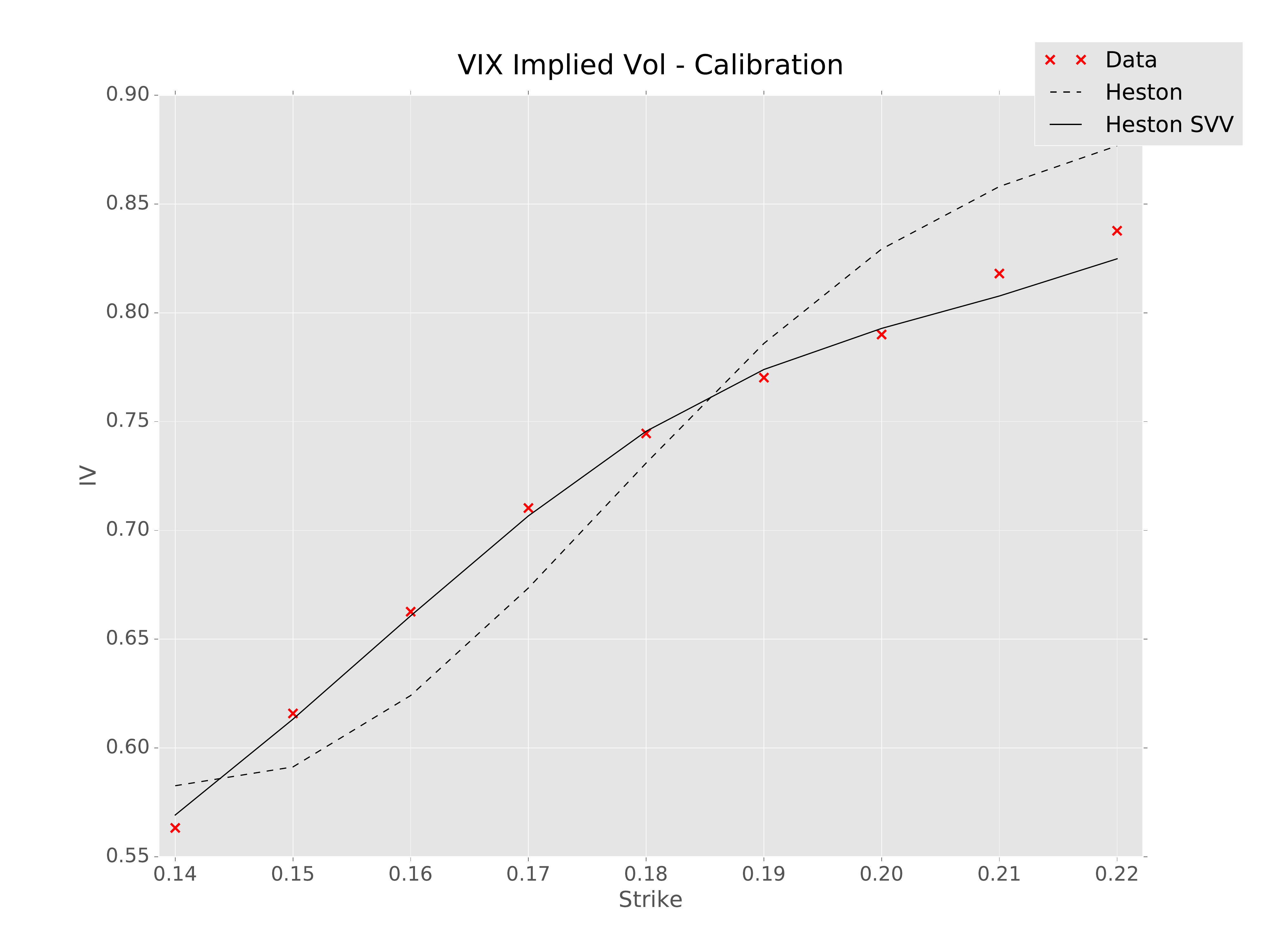}
    \vspace{-0.8cm}
    \caption{VIX}
    \label{fig:calibrated_iv_vix}
  \end{subfigure}
      \caption{Calibrated Implied Volatilities.}
      \label{fig:calibrated_iv}
\end{figure}

We present in Figure \ref{fig:calibrated_iv} the implied volatility and its calibrated approximation using the Heston and the Heston SVV models. One maturitiy of approximately 120 days was chosen in order to show the capacity of the model to capture both skews of the \spx\ and the VIX implied volatilities. Very short maturities are refrained because of the nature of our first-order approximation, see, for instance, \cite{multiscale_fouque_new_book}. As explained in Generalization \ref{gen:vix_time_dependent} in Section \ref{sec:generalization}, the term-structure of VIX could be captured by allowing time-dependence in some parameters of the model. 

\begin{figure}[h!]
  \begin{subfigure}{0.5\linewidth}
    \centering
    \includegraphics[width=\linewidth]{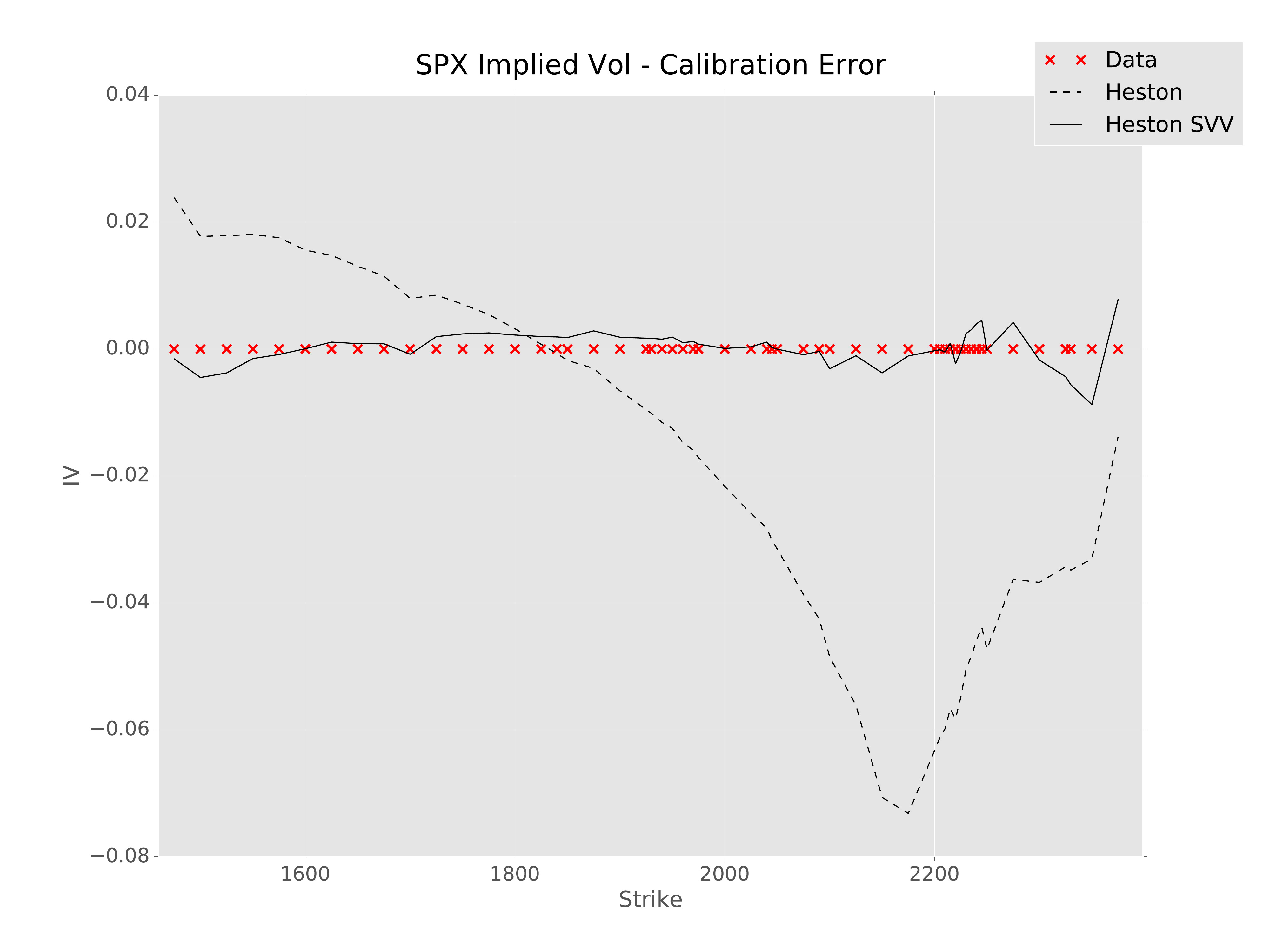}
    \vspace{-0.8cm}
    \caption{\spx}
    \label{fig:error_iv_spx}
  \end{subfigure}
  \begin{subfigure}{0.5\linewidth}
    \centering
    \includegraphics[width=\linewidth]{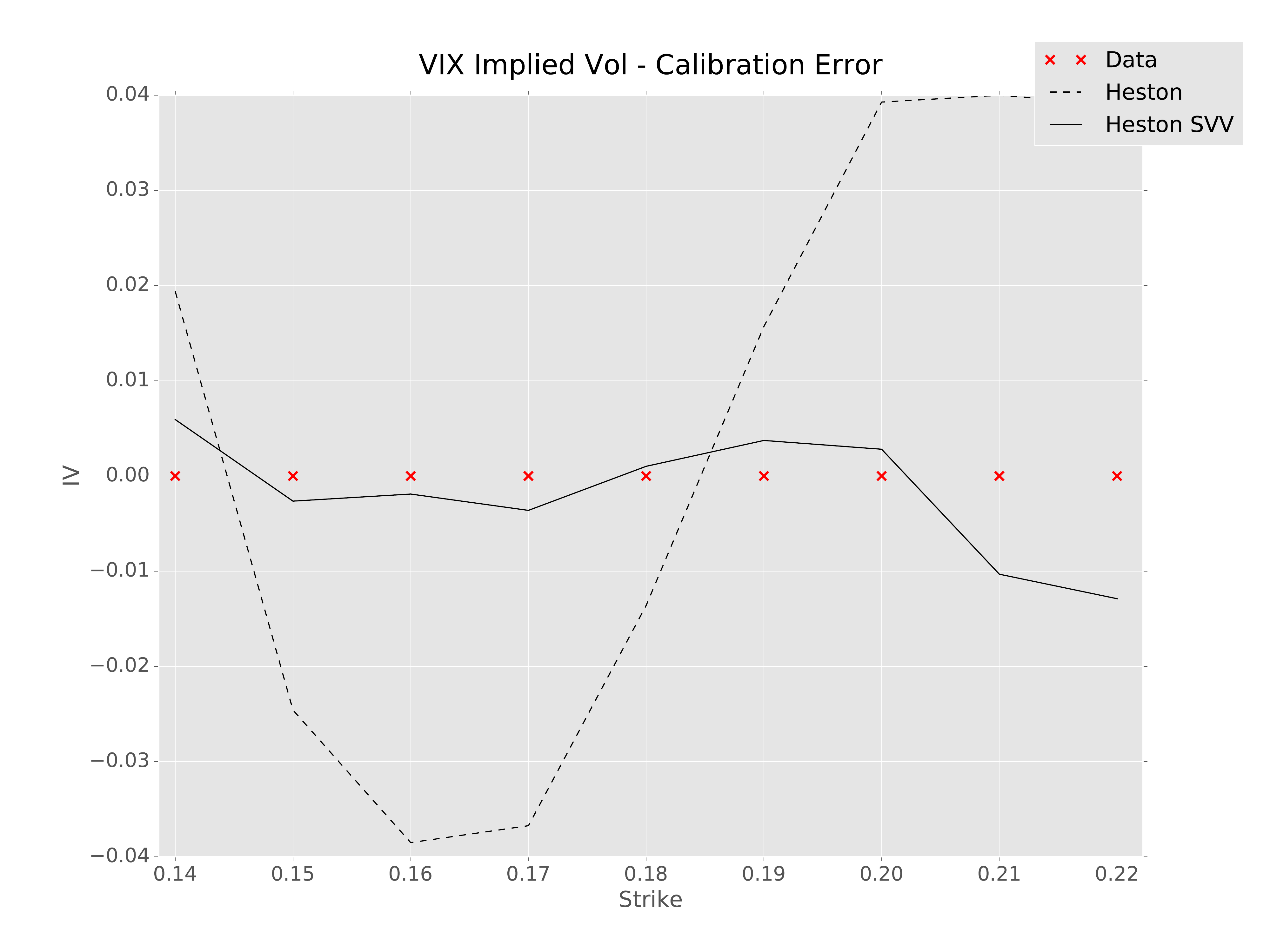}
    \vspace{-0.8cm}
    \caption{VIX}
    \label{fig:error_iv_vix}
  \end{subfigure}
      \caption{Calibration Errors.}
      \label{fig:error_iv}
\end{figure}

The calibrated parameters are shown in Table \ref{tab:calibrated}. The reader should notice that the $V^{\eps,\delta}$'s parameters satisfy their basic assumption of being small and that under the Heston SVV model, the vol-of-vol decreases compared to the standard Heston model, a desirable aspect for this model. In Table \ref{tab:mse} and Figure \ref{fig:error_iv}, one can observe a major improvement of the calibration when comparing the mean squared error, as in Equation (\ref{eq:calibration}). The enhancement of the calibration is better seen in \spx 's implied volatilities. Regarding VIX, Heston SVV model shows a better fit, presenting the concave shape of the curve.

\begin{filecontents*}{param_paper_20150821.csv}
Parameters,Heston,Heston  SVV
$\kappa$,17.38863,16.20866
$m$,0.04480,0.04275
$\eta$,3.70537,2.77650
$\rho$,-0.99000,-0.82897
,,
$V_{0 ,3}^{\eps}$,-,-0.00002
$V_{2 ,1}^{\eps}$,-,-0.00069
$V_{1 ,2}^{\eps}$,-,0.00243
$V_{1 ,0 ,\rho}^{\delta}$,-,0.05204
$V_{0 ,1 ,\rho}^{\delta}$,-,0.10119
$V_{1 ,0 ,\eta}^{\delta}$,-,-0.01469
$V_{0 ,1 ,\eta}^{\delta}$,-,-0.00180
\end{filecontents*}
\DTLloaddb{heston}{param_paper_20150821.csv}
\begin{filecontents*}{param_paper_error_20150821.csv}
$\empty$,MSE - Heston,MSE - Heston SVV
VIX,9.90067e-04,3.93575e-05
SPX,1.21396e-03,7.34144e-06
\end{filecontents*}
\DTLloaddb{error}{param_paper_error_20150821.csv}

\begin{table}[h!]
\centering
\renewcommand{\dtldisplayafterhead}{\hline  \hline }
\renewcommand{\dtldisplayendtab}{\tabularnewline \hline}
\renewcommand{\dtlstringalign}{c}
\renewcommand{\dtlrealalign}{c}
\DTLdisplaydb{heston}
\caption{Calibrated Parameters.}
\label{tab:calibrated}
\vspace{10pt}
\centering
\renewcommand{\dtldisplayafterhead}{\hline \hline}
\renewcommand{\dtldisplayendtab}{\tabularnewline \hline}
\renewcommand{\dtlstringalign}{c}
\renewcommand{\dtlrealalign}{c}
\DTLdisplaydb{error}
\caption{Mean Squared Error.}
\label{tab:mse}

\end{table}

\clearpage

\section{Generalizations}\label{sec:generalization}

Our model could be generalized in many ways, some simpler than others. Below we briefly discuss some of these possibilities:

\begin{generalization}

Notice that we could have explicitly considered the market prices of volatility risk as it is done in \cite{multiscale_fouque_new_book}, and in doing so we would have an additional term of order $\eps^{-1/2}$ and a term of order $\delta^{1/2}$ in the drifts of $Y^{\eps}$ and $Z^{\delta}$ respectively, both depending on $Y^{\eps}$ and $Z^{\delta}$. They could have been handled in the same way it is done in the aforesaid reference. For simplicity, we do not consider this generalization here.

\end{generalization}

\begin{generalization}

It is fairly simple to generalize the model above to deal with time-dependent interest rate and dividend yield. In order to make the exposition clearer, we will not consider it in the derivation of the first-order approximation shown in the sections to follow. However, this feature is present in the numerical computation in Section \ref{sec:numerical_example}.

\end{generalization}

\begin{generalization}\label{gen:vix_time_dependent}

In order to capture VIX's term structure, one would need to introduce time dependent parameters. The most parsimonious choice would be the long-run mean, $m$. This would add a moderate computational difficulty to our model, along with a more cumbersome notation and derivation of the first-order approximation, and therefore outside the scope of the paper. For a similar generalization, we refer the reader to \cite{sepp_vix_fourier} and \cite{heston_mikhailov}. Another approach to deal with VIX's term structure would be along the lines of the work \cite{Fouque2004}.

\end{generalization}

\begin{generalization}

It should be straight forward to adapt the machinery developed here to deal with the two-factor model of \cite{double_heston}:
\begin{align*}
\left\{
\begin{array}{l}
  dS_t = (r - q)S_t dt + \sqrt{V_t^{(1)}} S_t dW_t^{S^1} + \sqrt{V_t^{(2)}} S_t dW_t^{S^2}, \\
  dV_t^{(1)} = \kappa_1(m_1 - V_t^{(1)})dt + \eta_1(Y^{\eps}_t,Z^{\delta}_t) \sqrt{V_t^{(1)}} dW_t^{V^1},\\
  dV_t^{(2)} = \kappa_2(m_2 - V_t^{(2)})dt + \eta_2(Y^{\eps}_t,Z^{\delta}_t) \sqrt{V_t^{(2)}} dW_t^{V^2},
\end{array}
\right.
\end{align*}
with a simple correlation structure for the four-dimensional Brownian motion $(W_t^{S^1}, W_t^{S^2}, W_t^{V^1}, W_t^{V^2})$.

\end{generalization}

\begin{generalization}\label{gen:kappa}

A more complex generalization would be to consider that the mean-reverting rate, $\kappa$, is itself stochastic depending on $Y^\eps$ and $Z^\delta$ and that the long-run mean, $m$, satisfies $\kappa(y,z) m(y,z) = a$:
\begin{align*}
dV_t = (a  - \kappa(Y^{\eps}_t,Z^{\delta}_t)V_t)dt + \eta(Y^{\eps}_t,Z^{\delta}_t) \sqrt{V_t} dW_t^V.
\end{align*}

\end{generalization}

\section{Perturbation Framework}\label{sec:perturbation}

The first-order approximation for option prices on the the stock and its volatility index will be developed next. The arguments shown here justify the results and formulas presented in Section \ref{sec:results}. These arguments follow the ideas thoroughly explained in \cite{multiscale_fouque_new_book}.

\subsection{Options on the Stock}\label{sec:pertubation_S}

In this section, we derive the first-order approximation of the price of derivative contracts on $S$. Notice that, by the Feynman-Kac's Formula, $P_S^{\eps,\delta}$, that is given in Equation (\ref{eq:opt_price_S}), satisfies the following PDE
\begin{align}\label{eq:pde_S}
\left\{
\begin{array}{l}
  \cL^{\eps,\delta}_SP_S^{\eps,\delta}(t,s,v,y,z) = 0,  \\ \\
  P_S^{\eps,\delta}(T,s,v,y,z) = \varphir_S(s),
\end{array}
\right.
\end{align}
where the differential operator $\cL^{\eps,\delta}_S$ is given by
\begin{align*}
\cL^{\eps,\delta}_S &= \frac{1}{\eps} v\cL_0 + \frac{1}{\sqrt{\eps}} v \cL_1^S + \cL_2^S + \sqrt{\delta} v \cM_1^S + \delta v \cM_2 + \sqrt{\frac{\delta}{\eps}} v \cM_3,
\end{align*}
with
\begin{align}
\cL_0 &= \alpha(y) \frac{\partial}{\partial y} + \frac{1}{2} \beta^2(y) \frac{\partial^2}{\partial y^2}, \label{eq:cL_0} \\
\cL_1^S &= \rhor_{SY}  \beta(y) D_1 \frac{\partial}{\partial y} + \rhor_{VY} \eta(y,z)  \beta(y) \frac{\partial^2}{\partial v \partial y}, \label{eq:cL_1_S} \\
\cL_2^S &= \frac{\partial}{\partial t} + \frac{1}{2} v D_2 + (r-q)D_1 - r \cdot \label{eq:cL_2_S} \\
&+ \kappa(m - v)\frac{\partial}{\partial v} + \frac{1}{2} \eta^2(y,z) v\frac{\partial^2}{\partial v^2} + \rhor_{SV} \eta(y,z) v D_1 \frac{\partial}{\partial v}, \nonumber \\
\cM_1^S &= \rhor_{SZ}  g(z) D_1 \frac{\partial}{\partial z} + \rhor_{VZ} \eta(y,z) g(z) \frac{\partial^2}{\partial v \partial z}, \label{eq:cM_1_S}\\
\cM_2 &= c(z) \frac{\partial}{\partial z} + \frac{1}{2} g^2(z) \frac{\partial^2}{\partial z^2}, \label{eq:cM_2} \\
\cM_3 &= \rhor_{YZ}  \beta(y)g(z) \frac{\partial^2}{\partial y \partial z}, \label{eq:cM_3}\\
D_k &= s^k \frac{\partial^k}{\partial s^k}. \label{eq:Dk}
\end{align}

We now develop the singular and regular perturbation analysis for the option price $P_S^{\eps,\delta}$ following the method outlined in \cite{multiscale_fouque_new_book}. The reader will readily realize that the derivation is very similar to the Black--Scholes perturbation analysis performed in the aforesaid reference.

We formally write $P_S^{\eps,\delta}$ in powers of $\sqrt{\delta}$,
$$P_S^{\eps,\delta} = P^{\eps}_{S_0} + \sqrt{\delta} P^{\eps}_{S_1} +\delta P^{\eps}_{S_2} + \cdots,$$
and then, by Equation (\ref{eq:pde_S}), we choose $P^{\eps}_{S_0}$ and $P^{\eps}_{S_1}$ to satisfy
\begin{align}
&\left\{
\begin{array}{l}
  \ds \left(\frac{1}{\eps}v\cL_0 + \frac{1}{\sqrt{\eps}}v \cL_1^S + \cL_2^S\right)P^{\eps}_{S_0} = 0,  \\ \\
  P^{\eps}_{S_0}(T,s,v,y,z) = \varphir_S(s),
\end{array}
\right. \label{eq:pde_S0_eps}\\ \nonumber \\
&\left\{
\begin{array}{l}
  \ds\left(\frac{1}{\eps} v \cL_0 + \frac{1}{\sqrt{\eps}} v \cL_1^S + \cL_2^S\right)P^{\eps}_{S_1} = -\left( v \cM_1^S + \frac{1}{\sqrt{\eps}} v \cM_3 \right)P_{S_0}^{\eps},\\ \\
  P^{\eps}_{S_1}(T,s,v,y,z) = 0.
\end{array} \label{eq:pde_S1_eps}
\right.
\end{align}

\subsubsection{Computing $P_{S_0}$}\label{sec:com_P_S0}

We formally expand $P^{\eps}_{S_0}$ in powers of $\sqrt{\eps}$,
\begin{align}\label{eq:P_S_0_expansion}
P_{S_0}^{\eps} = \sum_{m \geq 0} (\sqrt{\eps})^m P_{S_{m,0}},
\end{align}
and denote $P_{S_{0,0}}$ simply by $P_{S_0}$, where we assume that, at maturity, $P_{S_0}(T,s,v,y,z) = \varphir_S(s)$, $P_{S_{1,0}}(T,s,v,y,z) = 0$ and $P_{S_{0,1}}(T,s,v,y,z) = 0$. Substituting expansion (\ref{eq:P_S_0_expansion}) into Equation (\ref{eq:pde_S0_eps}), we get the following PDEs:
\begin{align}
\hskip .5cm (-1,0):& \ v\cL_0P_{S_0} = 0, \label{eq:edp_eps_1}\\
\hskip .5cm (-1/2,0):& \ v\cL_0P_{S_{1,0}}+ v\cL_1^SP_{S_0} = 0, \label{eq:edp_eps_2}\\
\hskip .5cm (0,0):& \ v\cL_0P_{S_{2,0}}+ v\cL_1^SP_{S_{1,0}} + \cL_2^SP_{S_0}= 0, \label{eq:edp_eps_3}\\
\hskip .5cm (1/2,0):& \ v\cL_0P_{S_{3,0}} + v\cL_1^S P_{S_{2,0} }+ \cL_2^SP_{S_{1,0}} = 0, \label{eq:edp_eps_4}
\end{align}
with the notation $(i,j)$ denoting the term of $i$th order in $\eps$ and $j$th in $\delta$. Therefore, using the well-known arguments of \cite{multiscale_fouque_new_book}, we might choose:
\begin{itemize}

\item $P_{S_0} = P_{S_0}(t,s,v,z)$ and $P_{S_{1,0}} = P_{S_{1,0}}(t,s,v,z)$ independent of $y$;\\

\item $P_{S_0}$ satisfing $\langle \cL_2^S \rangle P_{S_0} = 0$;\\

\item $P_{S_{1,0}}$ solving $\langle \cL_2^S \rangle P_{S_{1,0}} = - v\langle \cL_1^S P_{S_{2,0}} \rangle$.\\

\end{itemize}

Define then the Heston differential operator:
\begin{align}
\cL_H(\eta,\rho) &= \cL_{BS}(\sqrt{v}) + \kappa(m - v)\frac{\partial}{\partial v} + \frac{1}{2} \eta^2 v \frac{\partial^2}{\partial v^2} + \rho \eta v D_1  \frac{\partial}{\partial v}, \label{eq:heston_op}
\end{align}
where $\cL_{BS}(\sigma)$ is the Black--Scholes differential operator with volatility $\sigma$:
\begin{align}
\cL_{BS}(\sigma) =  \frac{\partial}{\partial t} + \frac{1}{2} \sigma^2 s^2 \frac{\partial^2}{\partial s^2} + (r-q)s\frac{\partial}{\partial s} - r \cdot. \label{eq:bs_op}
\end{align}
Hence, by Equation (\ref{eq:cL_2_S}), $\cL_2^S = \cL_H(\eta(y,z),\rhor_{SV})$. Define now the averaged coefficients
\begin{align}
&\overline{\eta}(z) = \sqrt{\langle \eta^2(\cdot,z) \rangle}, \label{eq:eta_bar} \\
&\overline{\rho}(z) = \rhor_{SV} \frac{\langle \eta(\cdot,z) \rangle}{\overline{\eta}(z)}, \label{eq:rho_bar}
\end{align}
so that $\langle \cL_2^S \rangle = \cL_H(\overline{\eta}(z), \overline{\rho}(z))$ and $P_{S_0}$ solves the PDE
\begin{align}\label{eq:edp_p0}
\left\{
\begin{array}{l}
  \cL_H(\overline{\eta}(z), \overline{\rho}(z)) P_{S_0}(t,s,v,z) = 0,  \\ \\
  P_{S_0}(T,s,v,z) = \varphir_S(s).
\end{array}
\right.
\end{align}

The function $P_{S_0}$ might be computed using the method developed in \cite{heston93}, which is precisely described in Appendix \ref{sec:fourier_P_S0}.

\subsubsection{Computing $P^{\eps}_{S_{1,0}}$}

By the 0-order equation (\ref{eq:edp_eps_3}), the following formula holds true
\begin{align}
P_{S_{2,0}}(t,s,v,z) &= -\frac{1}{v}\cL_0^{-1}(\cL_2^S - \cL_H(\overline{\eta}(z), \overline{\rho}(z)))P_{S_0}(t,s,v,z) + c(t,s,v,z), \label{eq:P_S_20}
\end{align}
for some function $c$ that does not depend on $y$. Notice
\begin{align}
\cL_2^S - \cL_H(\overline{\eta}(z), \overline{\rho}(z))= &\frac{1}{2}(\eta^2(y,z) - \overline{\eta}^2(z)) v \frac{\partial^2}{\partial v^2} + \rhor_{SV} \left( \eta(y,z) - \langle \eta(\cdot,z) \rangle\right) v  D_1 \frac{\partial}{\partial v }.
\end{align}
Then, denote by $\phi(y,z)$ and $\psi(y,z)$ the solutions of the following Poisson equations
\begin{align}
&\cL_0 \phi(y,z) = \eta^2(y,z) - \overline{\eta}^2(z), \label{eq:poisson_phi} \\
&\cL_0 \psi(y,z) = \eta(y,z) - \langle \eta(\cdot,z) \rangle. \label{eq:poisson_psi}
\end{align}
Hence
\begin{align*}
\cL_0^{-1}(\cL_2^S  - \cL_H(\overline{\eta}(z), \overline{\rho}(z))) = \frac{1}{2} \phi(y,z) v \frac{\partial^2}{\partial v^2} + \rhor_{SV}  \psi(y,z) v D_1 \frac{\partial}{\partial v},
\end{align*}
and thus, Equation (\ref{eq:P_S_20}) implies
\begin{align*}
\cL_1^S P_{S_{2,0}} &= - \cL_1^S \left( \frac{1}{v} \cL_0^{-1}(\cL_2^S- \cL_H(\overline{\eta}(z), \overline{\rho}(z))P_{S_0}\right) \\
&= -\cL_1^S \left(\frac{1}{2} \phi(y,z)  \frac{\partial^2}{\partial v^2} + \rhor_{SV} \psi(y,z) D_1 \frac{\partial}{\partial v} \right)P_{S_0} \\
&= -\frac{1}{2} \rhor_{SY} \beta(y)  \frac{\partial \phi}{\partial y}(y,z) D_1 \frac{\partial^2 P_{S_0}}{\partial v^2} - \rhor_{SV} \rhor_{SY} \beta(y) \frac{\partial \psi}{\partial y}(y,z)  D_1^2 \frac{\partial P_{S_0}}{\partial v} \\
&- \frac{1}{2} \rhor_{VY} \eta(y,z) \beta(y) \frac{\partial \phi}{\partial y}(y,z) \frac{\partial^3 P_{S_0}}{\partial v^3} - \rhor_{SV} \rhor_{VY} \eta(y,z) \beta(y) \frac{\partial \psi}{\partial y}(y,z)  D_1 \frac{\partial^2 P_{S_0}}{\partial v^2}.
\end{align*}
Therefore, $P_{S_{1,0}}^{\eps} = \sqrt{\eps}P_{S_{1,0}}$ will be chosen to satisfy
\begin{align} \label{eq:edp_Peps_S_10}
\left\{
\begin{array}{l}
  \ds \cL_H(\overline{\eta}(z), \overline{\rho}(z))P^{\eps}_{S_{1,0}}(t,s,v,z) = -v \cA^{\eps} P_{S_0}(t,s,v,z),\\ \\
  P^{\eps}_{S_{1,0}}(T,s,v,z) = 0,
\end{array}
\right.
\end{align}
where
\begin{align}
&\cA^{\eps} = V_{1,2}^{\eps}(z) D_1 \frac{\partial^2}{\partial v^2} + V_{2,1}^{\eps}(z) D_1^2 \frac{\partial}{\partial v} +  V_{0,3}^{\eps}(z) \frac{\partial^3}{\partial v^3}, \\
&V_{1,2}^{\eps}(z) = -\sqrt{\eps}\frac{ \rhor_{SY} }{2} \left\langle \beta\frac{\partial \phi}{\partial y}(\cdot,z)  \right\rangle - \sqrt{\eps}\rhor_{SV} \rhor_{VY} \left\langle \eta(\cdot, z) \beta \frac{\partial \psi}{\partial y}(\cdot,z) \right\rangle, \label{eq:V12_eps}\\
&V_{2,1}^{\eps}(z) = - \sqrt{\eps}\rhor_{SV} \rhor_{SY} \left\langle \beta\frac{\partial \psi}{\partial y}(\cdot,z)  \right\rangle, \label{eq:V21_eps}\\
&V_{0,3}^{\eps}(z) = -\sqrt{\eps}\frac{ \rhor_{VY} }{2} \left\langle \eta(\cdot, z) \beta\frac{\partial \phi}{\partial y}(\cdot,z)  \right\rangle.\label{eq:V03_eps}
\end{align}

In Appendix \ref{sec:fourier_Peps_S10}, using Fourier transform techniques, we derive a quasi-closed formula for $P^{\eps}_{S_{1,0}}$.

\subsubsection{Computing $P^{\delta}_{S_{0,1}}$}\label{sec:comp_P01_delta}

We now expand $P_{S_1}^{\eps}$ in powers of $\sqrt{\eps}$,
$$P_{S_1}^{\eps} = \sum_{m \geq 0} (\sqrt{\eps})^m P_{S_{m,1}},$$
and then substitute this and the expansion for $P_{S_0}^{\eps}$ into Equation (\ref{eq:pde_S1_eps}) to find
\begin{align}
(-1,1/2):& \ v \cL_0 P_{S_{0,1}} = 0, \label{eq:edp_delta_1}\\
(-1/2,1/2):& \ v \cL_0 P_{S_{1,1}}+ v \cL_1^S P_{S_{0,1}} + v \cM_3 P_{S_0} = 0, \label{eq:edp_delta_2}\\
(0,1/2):& \ v\cL_0 P_{S_{2,1}}+ v \cL_1^S P_{S_{1,1}} + \cL_2^S P_{S_{0,1}} + v \cM_1^S P_{S_0} + v \cM_3P_{S_{1,0}} = 0. \label{eq:edp_delta_3}
\end{align}

Thus, we choose:

\begin{itemize}

\item $P_{S_{0,1}} = P_{S_{0,1}}(t,s,v,z)$ and $P_{S_{1,1}} = P_{S_{1,1}}(t,s,v,z)$ independent of $y$;\\

\item $P_{S_{0,1}}$ satisfying $\langle \cL_2^S \rangle P_{S_{0,1}} = -v \langle \cM_1^S \rangle P_{S_0}$.\\

\end{itemize}
Notice that
$$\langle \cM_1^S \rangle = \rhor_{SZ} g(z) D_1 \frac{\partial}{\partial z} + \rhor_{VZ} g(z) \langle \eta(\cdot,z) \rangle \frac{\partial^2}{\partial v \partial z},$$
and therefore,
\begin{align}\label{eq:edp_Pdelta_S_01}
\left\{
\begin{array}{l}
  \ds \cL_H(\overline{\eta}(z), \overline{\rho}(z)) P^{\delta}_{S_{0,1}}(t,s,v,z) = -v \cA^{\delta} P_{S_0}(t,s,v,z),\\ \\
  P^{\delta}_{S_{0,1}}(T,s,v,z)= 0,
\end{array}
\right.
\end{align}
where
\begin{align}
&\cA^{\delta} = V_{1,0,\eta}^{\delta}(z) D_1 \frac{\partial}{\partial \eta} + V_{1,0,\rho}^{\delta}(z) D_1 \frac{\partial}{\partial \rho} + V_{0,1,\eta}^{\delta}(z) \frac{\partial^2}{\partial v \partial \eta} + V_{0,1,\rho}^{\delta}(z) \frac{\partial^2}{\partial v \partial \rho},\\
&V_{0,1,\eta}^{\delta}(z) = \sqrt{\delta} \rhor_{VZ} g(z) \langle \eta(\cdot,z) \rangle \overline{\eta}'(z), \label{eq:V01eta_delta}\\
&V_{0,1,\rho}^{\delta}(z) = \sqrt{\delta} \rhor_{VZ} g(z) \langle \eta(\cdot,z) \rangle \overline{\rho}'(z), \label{eq:V01rho_delta}\\
&V_{1,0,\eta}^{\delta}(z) = \sqrt{\delta} \rhor_{SZ} g(z)\overline{\eta}'(z), \label{eq:V10eta_delta}\\
&V_{1,0,\rho}^{\delta}(z) = \sqrt{\delta} \rhor_{SZ} g(z)\overline{\rho}'(z).\label{eq:V10rho_delta}
\end{align}

In Appendix \ref{sec:fourier_Pdelta_S01}, using Fourier transform techniques, we find a quasi-closed formula for $P^{\delta}_{S_{0,1}}$.

\subsection{Options on the Volatility Index}\label{sec:pertubation_vix}

In this section, we will develop the first-order approximation for the price of options on VIX. Before continuing, it is necessary to study the dynamics of VIX under our model. Observe that, under mild conditions on $\eta(y,z)$, we have the well-known formula
\begin{align}
&\bE[V_u \ | \ \cF_t] = V_t e^{-\kappa(u - t)} + m(1 - e^{-\kappa(u -t)}),
\end{align}
which implies
\begin{align}\label{eq:vix_formula}
\VIX_t^2 &= \frac{1}{\tau_0} \int_t^{t+\tau_0}  \bE[V_u  \ | \ \cF_t ] du \\
&= m \left(1 - \frac{1 - e^{-\kappa \tau_0}}{\kappa \tau_0} \right) + \frac{1 - e^{-\kappa \tau_0}}{\kappa \tau_0}V_t = m(1-\theta) + \theta V_t, \nonumber
\end{align}
where
\begin{align}
& \theta = \frac{1 - e^{-\kappa \tau_0}}{\kappa \tau_0}.
\end{align}
Moreover, we define
\begin{align}
& \gamma(v) = \sqrt{m(1 - \theta) + \theta v}, \label{eq:g}
\end{align}
and notice that $\VIX_t = \gamma(V_t)$. This implies that we may consider, in this model, derivative contracts on VIX as contracts on $V$ but with a more complicated payoff. In Appendix \ref{sec:fourier_vix}, we present the Fourier method presented in \cite{sepp_vix_fourier} to compute the price of options on VIX, under constant vol-vol, using this observation.

Additionally, notice that, given $\VIX_t$, $m$ and $\kappa$, we can find the current value of $V$, $V_t$, using Equation (\ref{eq:vix_formula}).

\begin{remark}

Under the more complex model described in Generalization \ref{gen:kappa} in Section \ref{sec:generalization}, $\VIX_t$ would not be independent of $\eps$ and $\delta$. Indeed, it is fairly easy to show that
\begin{align}
\VIX_t^2 &= \frac{1}{\tau_0} \int_t^{t+\tau_0}  \bE[V_u  \ | \ \cF_t ] du \\
&= F_{0,0}(\tau_0,z) + vF_{0,1}(\tau_0,z) + V_{1,\kappa}^{\delta}(z)(F_{1,0}(\tau_0,z) + vF_{1,1}(\tau_0,z)) + O(\eps + \delta),
\end{align}
for some functions $F_{i,j}$ that could be explicitly computed. The constant $V_{1,\kappa}^{\delta}(z)$ is related to the fast and slow time scales in $\kappa$. Therefore, in order to compute the first-order approximation for option on VIX, one should consider the same approach as in \cite{mutiscale_saporito_future}. In this paper, the authors used the first-order approximation of future prices and examined the problem of computing the first-order approximation on derivatives on futures as a singular and regular perturbation of an asset whose dynamics itself is only known up to its first-order approximation.

\end{remark}

We will now derive the first-order approximation for derivatives contracts on VIX. Define the following differential operator
\begin{align*}
\cL^{\eps,\delta}_V = \frac{1}{\eps} v \cL_0 + \frac{1}{\sqrt{\eps}} v \cL_1^V + \cL_2^V + \sqrt{\delta}v  \cM_1^V + \delta v \cM_2 + \sqrt{\frac{\delta}{\eps}} v \cM_3,
\end{align*}
where $\cL_0$, $\cM_2$ and $\cM_3$ are the same as in the pricing PDE for derivatives on $S$, see Equations (\ref{eq:cL_0}), (\ref{eq:cM_2}) and (\ref{eq:cM_3}), respectively, and $\cL_1^V$, $\cL_2^V$ and $\cM_1^V$ are given by
\begin{align}
&\cL_1^V = \rhor_{VY}  \eta(y,z)\beta(y) \frac{\partial^2}{\partial v \partial y}, \\
&\cL_2^V = \frac{\partial}{\partial t} + \kappa(m - v) \frac{\partial}{\partial v} + \frac{1}{2}\eta^2(y,z) v \frac{\partial^2}{\partial v^2} - r\cdot, \\
&\cM_1^V =\rhor_{VZ} \eta(y,z)g(z) \frac{\partial^2}{\partial v \partial z}.
\end{align}
Hence, $P^{\eps,\delta}_V$, defined in Equation (\ref{eq:opt_price_VIX}), satisfies the following PDE:
$$\left\{
\begin{array}{l}
  \cL^{\eps,\delta}_VP^{\eps,\delta}_V(t,v,y,z) = 0,  \\ \\
  P^{\eps,\delta}_V(T,v,y,z) = \varphir_V(\gamma(v)),
\end{array}
\right.$$
where $\gamma$ is given by Equation (\ref{eq:g}). Following \cite{multiscale_fouque_new_book}, as we have done in the previous section, we conclude that the first-order approximation for $P^{\eps,\delta}_V$ solves the following PDEs
\begin{align}
&\left\{
\begin{array}{l}
  \cLcir (\overline{\eta}(z))P_{V_0}(t,v,z) = 0,  \\ \\
  P_{V_0}(T,v,z) = \varphir_V(\gamma(v)),
\end{array}
\right.\label{eq:P0_vix}\\ \nonumber \\
&\left\{
\begin{array}{l}
  \ds \cLcir (\overline{\eta}(z))P^{\eps}_{V_{1,0}}(t,v,z) = -v V_3^{\eps}(z) \frac{\partial^3 P_{V_0}}{\partial v^3}(t,v,z)  \\ \\
  P^{\eps}_{V_{1,0}}(T,v,z) = 0,
\end{array}
\right. \label{eq:P10eps_vix}\\ \nonumber \\
&\left\{
\begin{array}{l}
  \ds \cLcir (\overline{\eta}(z))P^{\delta}_{V_{0,1}}(t,v,z) = - v V_1^{\delta}(z) \frac{\partial^2 P_{V_0}}{\partial v \partial \eta}(t,v,z),  \\ \\
  P^{\delta}_{V_{0,1}}(T,v,z) = 0,
\end{array}
\right. \label{eq:P01delta_vix}
\end{align}
where
\begin{align}
\cLcir (\eta) &= \frac{\partial}{\partial t} + \kappa(m - v) \frac{\partial}{\partial v} + \frac{1}{2} \eta^2 v\frac{\partial^2}{\partial v^2} - r\cdot, \label{eq:cL_CIR}\\
V_3^{\eps}(z) &= -\sqrt{\eps}\frac{ \rhor_{VY} }{2} \left\langle \frac{\partial \phi}{\partial y}(\cdot,z) \eta(\cdot,z)\beta \right\rangle, \label{eq:V3_eps_vix}\\
V_1^{\delta}(z) &= \sqrt{\delta} \rhor_{VZ} g(z)  \langle \eta(\cdot,z) \rangle \overline{\eta}'(z). \label{eq:V1delta_vix}
\end{align}
We are using the notation $\cLcir (\eta)$ because it is related to the infinitesimal generator of a CIR process with constant volatility $\eta$. The quasi-closed formulas for $P_{V_0}$, $P^{\eps}_{V_{1,0}}$ and $P^{\delta}_{V_{0,1}}$ are given in Appendix \ref{sec:fourier_vix}.

\subsection{Accuracy of the Approximation}\label{sec:accuracy}

We now state the precise accuracy result for the formal approximation determined in the previous sections. All the reasoning in Sections \ref{sec:pertubation_S} and \ref{sec:pertubation_vix} were only formal arguments and well-thought \textit{choices} for the proposed first-order approximations. The following theorem is the result that establishes the order of accuracy of this approximation and justifies, a posteriori, the choices made earlier. The proof is very similar to the ones presented in \cite{multiscale_fouque_new_book} and \cite{fouque_fast_heston} and, therefore, omitted.

\begin{theorem}\label{thm:accuracy_teo}

Under Assumption \ref{ass:model} and if the payoff functions $\varphir_S$ and $\varphir_V$ are continuous and piecewise smooth, then
\begin{align*}
P^{\eps,\delta}_S(t,s,v,y,z) &= P_{S_0}(t,s,v,z) + P^{\eps}_{S_{1,0}}(t,s,v,z) + P^{\delta}_{S_{0,1}}(t,s,v,z) + O(\eps + \delta),\\
P^{\eps,\delta}_V(t,v,y,z) &= P_{V_0}(t,v,z) + P^{\eps}_{V_{1,0}}(t,v,z) + P^{\delta}_{V_{0,1}}(t,v,z) + O(\eps + \delta).
\end{align*}

\end{theorem}

\section{Conclusion}\label{sec:conclusion}

In this paper, we have proposed a continuous diffusion model for the stock price that is able to capture both skews in the stock's and volatility index's options data and that allows for quasi-closed formulas for the first-order approximation for option prices on the spot and its volatility index. These features were not achieve by any other continuous diffusion model. We have exemplified our calibration procedure with real data on \spx\ and VIX. 

Further research could be conducted in order to develop the generalizations outlined in Section \ref{sec:generalization}. For instance, to be able to achieve the fit of VIX's term structure, one could consider the method outlined in this paper to derive the first-order approximation under a time-dependent generalization of the model proposed here.

\appendices

\section{Fourier Method to Compute the First-Order Approximation for Options on $S$}\label{sec:fourier_S}

The computations presented here are based on the ideas shown in \cite{fouque_fast_heston}.\\

Let us first change variables to better apply the Fourier method.
\begin{align}
&\tau(t) = T - t, \\
&x(t,s) = (r-q)(T-t) + \log s, \\
&\widetilde{P}_{S_0}(\tau,x,v,z) = e^{r\tau}P_{S_0}(T-\tau,e^{x - (r-q)\tau},v,z), \\
&\widetilde{P}_{S_{1,0}}^{\eps}(\tau,x,v,z) = e^{r\tau}P^{\eps}_{S_{1,0}}(T-\tau,e^{x - (r-q)\tau},v,z), \\
&\widetilde{P}_{S_{0,1}}^{\delta}(\tau,x,v,z) = e^{r\tau}P^{\delta}_{S_{0,1}}(T-\tau,e^{x - (r-q)\tau},v,z), \\
&\widetilde{\cL}_H = -\frac{\partial}{\partial \tau} + \frac{1}{2} v \left( \frac{\partial^2}{\partial x^2} - \frac{\partial}{\partial x}\right) + \kappa(m - v)\frac{\partial}{\partial v} + \frac{1}{2} \overline{\eta}(z)^2 v \frac{\partial^2}{\partial v^2} + \overline{\rho}(z)\overline{\eta}(z)v \frac{\partial^2}{\partial x \partial v},\\
&\widetilde{\cA^{\eps}} = V_{1,2}^{\eps}(z) \frac{\partial}{\partial x} \frac{\partial^2}{\partial v^2} + V_{2,1}^{\eps}(z) \frac{\partial^2}{\partial x^2} \frac{\partial}{\partial v} +  V_{0,3}^{\eps}(z) \frac{\partial^3}{\partial v^3},\\
&\widetilde{\cA^{\delta}} = V_{1,0,\eta}^{\delta}(z) \frac{\partial^2}{\partial \eta \partial x} + V_{1,0,\rho}^{\delta}(z)  \frac{\partial^2}{\partial \rho \partial x} + V_{0,1,\eta}^{\delta}(z) \frac{\partial^2}{\partial v \partial \eta} + V_{0,1,\rho}^{\delta}(z) \frac{\partial^2}{\partial v \partial \rho}.
\end{align}

Therefore, one concludes
\begin{align}
&\left\{
\begin{array}{l}
  \ds \widetilde{\cL}_H  \widetilde{P}_{S_0}(\tau,x,v,z) =0,\\ \\
  \widetilde{P}_{S_0}(0,x,v,z)= \widetilde{\varphir_S}(x),
\end{array}
\right. \label{eq:edp_P_S_0_tilde} \\ \nonumber \\
&\left\{
\begin{array}{l}
  \ds \widetilde{\cL}_H  \widetilde{P}_{S_{1,0}}^{\eps}(\tau,x,v,z) = -v \widetilde{\cA^{\eps}} \widetilde{P}_{S_0}(\tau,x,v,z),\\ \\
   \widetilde{P}_{S_{1,0}}^{\eps}(0,x,v,z)= 0,
\end{array}
\right. \label{eq:edp_Peps_S_10_tilde} \\ \nonumber \\
&\left\{
\begin{array}{l}
  \ds \widetilde{\cL}_H  \widetilde{P}_{S_{0,1}}^{\delta}(\tau,x,v,z) = -v \widetilde{\cA^{\delta}} \widetilde{P}_{S_0}(\tau,x,v,z),\\ \\
   \widetilde{P}_{S_{0,1}}^{\delta}(0,x,v,z)= 0.
\end{array}
\right. \label{eq:edp_Pdelta_S_01_tilde}
\end{align}

\subsection{A Quasi-Closed Formula for $P_{S_0}$}\label{sec:fourier_P_S0}

Define
\begin{align}
&\widehat{\cL}_H = -\frac{\partial}{\partial \tau} + \frac{1}{2} v (-\xi^2 + i\xi) + (\kappa m - (\kappa +  \overline{\rho}(z)\overline{\eta}(z) i\xi)v)\frac{\partial}{\partial v} + \frac{1}{2}
\overline{\eta}^2(z)v \frac{\partial^2}{\partial v^2}.
\end{align}
Hence, if we denote by $\widehat{P}_{S_0}(\tau,\xi,v,z)$ the Fourier transform of $\widetilde{P}_{S_0}(\tau,x,v,z)$ with respect to $x$, $\widehat{P}_{S_0}$ satisfies the following PDE
\begin{align}
&\left\{
\begin{array}{l}
  \ds \widehat{\cL}_H \widehat{P}_{S_0}(\tau,\xi,v,z) =0,\\ \\
  \widehat{P}_{S_0}(0,\xi,v,z)= \widehat{\varphir_S}(\xi).
\end{array}
\right.
\end{align}
Therefore, by the arguments presented in \cite{heston93}, we can write
\begin{align}
P_{S_0}(t,s,v,z) &= \frac{e^{-r \tau}}{\pi} \int_0^{+\infty} Re\left( e^{-i\xi x(t,s)} G_S(\tau,\xi,v,z)\widehat{\varphir_S}(\xi) \right) d\xi_r,
\end{align}
where
\begin{align}
&\xi = \xi_r + i\xi_i,\\
&G_S(\tau,\xi,v,z) = e^{C(\tau,\xi,z) + vD(\tau,\xi,z)}, \\
&C(\tau,\xi,z) = \frac{\kappa m }{\overline{\eta}^2(z)} \left( (\kappa + i\overline{\rho}(z) \overline{\eta}(z)\xi - d(\xi, z)) \tau - 2 \log\left( \frac{e^{-d(\xi, z)\tau}/g(\xi, z) - 1}{1/g(\xi, z) - 1} \right) \right), \\
&D(\tau,\xi,z) = \frac{\kappa + i\overline{\rho}(z) \overline{\eta}(z)\xi + d(\xi, z)}{\overline{\eta}^2(z)}\left( \frac{1 - e^{d(\xi, z)\tau}}{1 - g(\xi, z)e^{d(\xi, z)\tau}} \right), \\
&g(\xi, z) = \frac{\kappa + i\overline{\rho}(z) \overline{\eta}(z)\xi + d(\xi, z)}{\kappa + i\overline{\rho}(z) \overline{\eta}(z)\xi - d(\xi, z)},\\
&d(\xi, z) = \sqrt{\overline{\eta}^2(z)(\xi^2 - i\xi) + (\kappa + i\overline{\rho}(z) \overline{\eta}(z)\xi)^2}.
\end{align}
For call options, we must set $\xi_i > 1$.

\subsection{A Quasi-Closed Formula for $P^{\eps}_{S_{1,0}}$}\label{sec:fourier_Peps_S10}

If we denote by $\widehat{P}_{S_{1,0}}^{\eps}(\tau,\xi,v,z)$ the Fourier transform of $\widetilde{P}_{S_{1,0}}^{\eps}(\tau,x,v,z)$ with respect to $x$, $\widehat{P}_{S_{1,0}}^{\eps}$ satisfies the following PDE
$$\widehat{\cL}_H \widehat{P}_{S_{1,0}}^{\eps}(\tau,\xi,v,z) = -v\widehat{\cA^{\eps}} G_S(\tau,\xi,v,z)\widehat{\varphir_S}(\xi),$$
where
\begin{align}
&\widehat{\cA^{\eps}} = -i\xi V_{1,2}^{\eps}(z)\frac{\partial^2}{\partial v^2} -\xi^2 V_{2,1}^{\eps}(z) \frac{\partial}{\partial v} +  V_{0,3}^{\eps}(z) \frac{\partial^3}{\partial v^3}.
\end{align}
Consider now the following \textit{ansatz}:
$$\widehat{P}_{S_{1,0}}^{\eps}(\tau,\xi,v,z) = (f^{\eps,\delta}_0(\tau,\xi,z) + vf^{\eps,\delta}_1(\tau,\xi,z)) G_S(\tau,\xi,v,z)\widehat{\varphir_S}(\xi).$$
So,
\begin{align*}
&\widehat{\cA^{\eps}} G_S = \left(-i\xi V_{1,2}^{\eps}(z)D^2 -\xi^2 V_{2,1}^{\eps}(z)  D + V_{0,3}^{\eps}(z) D^3\right) G_S,\\
&\widehat{\cL}_H \widehat{P}_{S_{1,0}}^{\eps} = \left(-\frac{\partial f^{\eps}_0}{\partial \tau} -v\frac{\partial f^{\eps}_1}{\partial \tau} \right)G_S\widehat{\varphir_S} - (f^{\eps}_0 + vf^{\eps}_1) \frac{\partial G_S}{\partial \tau}\widehat{\varphir_S}  \\
&+ (\kappa m - (\kappa +  \overline{\rho}(z)\overline{\eta}(z) i\xi)v) \left(f^{\eps}_1 G_S\widehat{\varphir_S} + (f^{\eps}_0 + vf^{\eps}_1) \frac{\partial G_S}{\partial v}\widehat{\varphir_S}\right) \\
& + \frac{1}{2} v (-\xi^2 + i\xi)(f^{\eps}_0 + vf^{\eps}_1) G_S\widehat{\varphir_S} + \frac{1}{2} \overline{\eta}^2(z)v \left(2f^{\eps}_1\frac{\partial G_S}{\partial v}\widehat{\varphir_S} + (f^{\eps}_0 + vf^{\eps}_1) \frac{\partial^2 G_S}{\partial v^2}\widehat{\varphir_S} \right) \\
&= \cancelto{\scriptstyle 0}{\left(\widehat{\cL}_H G_S \right)}(f^{\eps}_0 + vf^{\eps}_1)\widehat{\varphir_S} + \left(-\frac{\partial f^{\eps}_0}{\partial \tau} -v\frac{\partial f^{\eps}_1}{\partial \tau} \right)\overline{G}_S\widehat{\varphir_S} \\
&+ (\kappa m - (\kappa +  \overline{\rho}(z)\overline{\eta}(z) i\xi)v)f^{\eps}_1 G_S\widehat{\varphir_S} + \overline{\eta}^2(z)vf^{\eps}_1 D G_S\widehat{\varphir_S} \\
&= \left(-\frac{\partial f^{\eps}_0}{\partial \tau} -v\frac{\partial f^{\eps}_1}{\partial \tau} + (\kappa m - (\kappa +  \overline{\rho}(z)\overline{\eta}(z) i\xi)v)f^{\eps}_1 +  \overline{\eta}^2(z)vf^{\eps}_1 D \right)G_S\widehat{\varphir_S},
\end{align*}
and hence we should choose $f^{\eps}_0$ and $f^{\eps}_1$ to solve
\begin{align}\label{eq:ode_system_P_S_eps}
\left\{
\begin{array}{l}
\ds \frac{\partial f^{\eps}_1}{\partial \tau}(\tau,\xi,z) =  (\overline{\eta}^2(z)D(\tau,\xi,z) - (\kappa +  \overline{\rho}(z)\overline{\eta}(z) i\xi))f^{\eps}_1(\tau,\xi,z) \\ \\
 \ds \hskip 1cm -i\xi V_{1,2}^{\eps}(z)D^2(\tau,\xi,z) -\xi^2 V_{2,1}^{\eps}(z) D(\tau,\xi,z) +  V_{0,3}^{\eps}(z)D^3(\tau,\xi,z), \\ \\
\ds \frac{\partial f^{\eps}_0}{\partial \tau}(\tau,\xi,z) = \kappa m f^{\eps}_1(\tau,\xi,z),  \\ \\
f^{\eps}_0(0,\xi,z) = f^{\eps}_1(0,\xi,z) = 0.
\end{array}
\right.
\end{align}
Therefore, once we solve the ODE system above, we need to compute
\begin{align}
P^{\eps}_{S_{1,0}}(t,s,v,z) = \frac{e^{-r \tau}}{\pi} \int_0^{+\infty} Re\left( e^{-i\xi x(t,s)} (f^{\eps}_0 + vf^{\eps}_1) G_S(\tau,\xi,v,z)\widehat{\varphir_S}(\xi) \right) d\xi_r.
\end{align}

\subsection{A Quasi-Closed Formula for $P^{\delta}_{S_{0,1}}$}\label{sec:fourier_Pdelta_S01}

Now, if we denote by $\widehat{P}_{S_{0,1}}^{\delta}(\tau,\xi,v,z)$ the Fourier transform of $\widetilde{P}_{S_{0,1}}^{\delta}(\tau,x,v,z)$ with respect to $x$, $\widehat{P}_{S_{0,1}}^{\delta}$ satisfies the following PDE
$$\widehat{\cL}_H \widehat{P}_{S_{0,1}}^{\delta}(\tau,\xi,v,z) = -v\widehat{\cA^{\delta}} G_S(\tau,\xi,v,z)\widehat{\varphir_S}(\xi),$$
where
$$\widehat{\cA^{\delta}} = -i\xi V_{1,0,\eta}^{\delta}(z) \frac{\partial}{\partial \eta} -i\xi V_{1,0,\rho}^{\delta}(z)  \frac{\partial}{\partial \rho} + V_{0,1,\eta}^{\delta}(z) \frac{\partial^2}{\partial v \partial \eta} + V_{0,1,\rho}^{\delta}(z) \frac{\partial^2}{\partial v \partial \rho}.$$
Consider now the following \textit{ansatz}:
$$\widehat{P}_{S_{0,1}}^{\delta}(\tau,\xi,v,z) = ( g^{\delta}_0(\tau,\xi,z) + v g^{\delta}_1(\tau,\xi,z) + v^2 g^{\delta}_2(\tau,\xi,z)) G_S(\tau,\xi,v,z)\widehat{\varphir_S}(\xi),$$
and notice
\begin{align*}
&\widehat{\cA^{\delta}} G_S(\tau,\xi,v,z) = \left(-i\xi V_{1,0,\eta}^{\delta}(z) \left( \frac{\partial C}{\partial \eta} + v \frac{\partial D}{\partial \eta}\right) -i\xi V_{1,0,\rho}^{\delta}(z) \left( \frac{\partial C}{\partial \rho} + v \frac{\partial D}{\partial \rho}\right) \right. \\
&\left. + V_{0,1,\eta}^{\delta}(z) \left( \frac{\partial C}{\partial \eta} + \frac{\partial D}{\partial \eta} + v \frac{\partial D}{\partial \eta}\right) + V_{0,1,\rho}^{\delta}(z) \left( \frac{\partial C}{\partial \rho} + \frac{\partial D}{\partial \rho} + v \frac{\partial D}{\partial \rho}\right)\right) G_S(\tau,\xi,v,z) \\
&= \left((V_{0,1,\eta}^{\delta}(z)-i\xi V_{1,0,\eta}^{\delta}(z)) \frac{\partial C}{\partial \eta} + (V_{0,1,\rho}^{\delta}(z) -i\xi V_{1,0,\rho}^{\delta}(z)) \frac{\partial C}{\partial \rho}  \right.\\
& + V_{0,1,\eta}^{\delta}(z) \frac{\partial D}{\partial \eta} + \left.V_{0,1,\rho}^{\delta}(z) \frac{\partial D}{\partial \rho} \right) G_S(\tau,\xi,v,z) \\
&+ v \left((V_{0,1,\eta}^{\delta}(z) -i\xi V_{1,0,\eta}^{\delta}(z) )\frac{\partial D}{\partial \eta} + (V_{0,1,\rho}^{\delta}(z)\ -i\xi V_{1,0,\rho}^{\delta}(z) )\frac{\partial D}{\partial \rho}\right) G_S(\tau,\xi,v,z).
\end{align*}
Therefore,
\begin{align*}
&\widehat{\cL}_H \widehat{P}_{S_{0,1}}^{\delta} = \left(-\frac{\partial g^{\delta}_0}{\partial \tau} -v\frac{\partial g^{\delta}_1}{\partial \tau} -v^2\frac{\partial g^{\delta}_2}{\partial \tau}\right)G_S\widehat{\varphir_S} - ( g^{\delta}_0 + v g^{\delta}_1 + v^2 g^{\delta}_2) \frac{\partial G_S}{\partial \tau}\widehat{\varphir_S}\\
&+ \frac{1}{2} v (-\xi^2 + i\xi)( g^{\delta}_0 + v g^{\delta}_1 + v^2 g^{\delta}_2) G_S\widehat{\varphir_S} \\
&+ (\kappa m - (\kappa +  \overline{\rho}(z)\overline{\eta}(z) i\xi)v) \left( g^{\delta}_1 G_S\widehat{\varphir_S} + 2v g^{\delta}_2 G_S\widehat{\varphir_S} + ( g^{\delta}_0 + v g^{\delta}_1 + v^2 g^{\delta}_2) \frac{\partial G_S}{\partial v}\widehat{\varphir_S}\right)\\
&+ \frac{1}{2} \overline{\eta}^2(z)v \left(2 g^{\delta}_1 \frac{\partial G_S}{\partial v}\widehat{\varphir_S} + 2 g^{\delta}_2 G_S\widehat{\varphir_S} + 4v g^{\delta}_2 \frac{\partial G_S}{\partial v}\widehat{\varphir_S} + ( g^{\delta}_0 + v g^{\delta}_1 + v^2 g^{\delta}_2) \frac{\partial^2 G_S}{\partial v^2}\widehat{\varphir_S}\right) \\
&= \cancelto{\scriptstyle 0}{\left(\widehat{\cL}_H G_S\right)}( g^{\delta}_0 + v g^{\delta}_1 + v^2 g^{\delta}_2)\widehat{\varphir_S} + \left(-\frac{\partial g^{\delta}_0}{\partial \tau} -v\frac{\partial g^{\delta}_1}{\partial \tau} -v^2\frac{\partial g^{\delta}_2}{\partial \tau}\right)G_S\widehat{\varphir_S} \\
&+ (\kappa m - (\kappa +  \overline{\rho}(z)\overline{\eta}(z) i\xi)v)\left( g^{\delta}_1 + 2v g^{\delta}_2\right) G_S\widehat{\varphir_S} + \overline{\eta}^2(z)v( g^{\delta}_1 D +  g^{\delta}_2 + 2v g^{\delta}_2 D) G_S\widehat{\varphir_S}\\
&= \left(-\frac{\partial g^{\delta}_0}{\partial \tau} -v\frac{\partial g^{\delta}_1}{\partial \tau} -v^2\frac{\partial g^{\delta}_2}{\partial \tau} + (\kappa m - (\kappa +  \overline{\rho}(z)\overline{\eta}(z) i\xi)v)\left( g^{\delta}_1 + 2v g^{\delta}_2\right) \right.  \\
&+ \left. \overline{\eta}^2(z)v( g^{\delta}_1 D +  g^{\delta}_2 + 2v g^{\delta}_2 D) \right)G_S\widehat{\varphir_S},
\end{align*}
and hence we should choose $ g^{\delta}_0$, $ g^{\delta}_1$ and $ g^{\delta}_2$ to solve
\begin{align}\label{eq:ode_system_P_S_delta}
\left\{\begin{array}{l}
\ds \frac{\partial g^{\delta}_2}{\partial \tau}(\tau,\xi,z) = -2(\kappa +  \overline{\rho}(z)\overline{\eta}(z) i\xi - \overline{\eta}^2(z) D(\tau,\xi,z)) g^{\delta}_2(\tau,\xi,z) \\ \\
\hskip 1cm \ds (V_{0,1,\eta}^{\delta}(z) -i\xi V_{1,0,\eta}^{\delta}(z) )\frac{\partial D}{\partial \eta} + (V_{0,1,\rho}^{\delta}(z)\ -i\xi V_{1,0,\rho}^{\delta}(z) )\frac{\partial D}{\partial \rho}, \\ \\
\ds \frac{\partial g^{\delta}_1}{\partial \tau}(\tau,\xi,z) = -(\kappa +  \overline{\rho}(z)\overline{\eta}(z) i\xi - \overline{\eta}^2(z) D(\tau,\xi,z)) g^{\delta}_1(\tau,\xi,z) \\ \\
 \ds \hskip 1cm + \overline{\eta}^2(z) g^{\delta}_2(\tau,\xi,z) + (V_{0,1,\eta}^{\delta}(z)-i\xi V_{1,0,\eta}^{\delta}(z)) \frac{\partial C}{\partial \eta} + (V_{0,1,\rho}^{\delta}(z) -i\xi V_{1,0,\rho}^{\delta}(z)) \frac{\partial \overline{C}}{\partial \rho},  \\ \\
\ds \hskip 1cm + V_{0,1,\eta}^{\delta}(z) \frac{\partial D}{\partial \eta} + V_{0,1,\rho}^{\delta}(z) \frac{\partial D}{\partial \rho} \\ \\
\ds \frac{\partial g^{\delta}_0}{\partial \tau}(\tau,\xi,z) = \kappa m  g^{\delta}_1(\tau,\xi,z), \\ \\
 g^{\delta}_0(0,\xi,z) =  g^{\delta}_1(0,\xi,z) =  g^{\delta}_2(0,\xi,z) = 0.
\end{array}
\right.
\end{align}
Therefore, once we solve the ODE system above, we need to compute
\begin{align}
P^{\delta}_{S_{0,1}}(t,s,v,z) =  \frac{e^{-r \tau}}{\pi}\int_0^{+\infty} Re\left(e^{-i\xi x(t,s)} ( g^{\delta}_0 + v g^{\delta}_1 + v^2  g^{\delta}_2) G_S(\tau,\xi,v,z)\widehat{\varphir_S}(\xi)\right)d\xi_r.
\end{align}

\section{Fourier Method to Compute the First-Order Approximation for Options on VIX}\label{sec:fourier_vix}

Firstly, notice that
$$P_{V_0}(t,v,z) = \bE[e^{-r(T-t)}\varphir_V(g(\overline{V}_T)) \ | \ \overline{V}_t  = v],$$
where $d\overline{V}_t = \kappa(m - \overline{V}_t)dt + \overline{\eta}(z) \sqrt{\overline{V}_t} dW_t^V$. Hence,
$$\left\{
\begin{array}{l}
\cLcir(\overline{\eta}(z)) P_{V_0}(t,v,z) = 0, \\ \\
P_{V_0}(T,v,z) = \varphir_V(\gamma(v)).
\end{array}
\right.$$
Using the Green's function technique, as in \cite{sepp_vix_fourier}, the solution of this PDE can be written as
\begin{align}
P_{V_0}(t,v,z) = \frac{e^{-r\tau}}{\pi} \int_0^{+\infty}G_V(t,v,\nu, z) \widehat{\varphir_V}(\nu) d\nu_i, \label{eq:P0_vix_formula}
\end{align}
where $\tau = T-t$, $\nu = \nu_r + i \nu_i$, $\nu_r \leq 0$ and
\begin{align*}
&\widehat{\varphir_V}(\nu) = \int_0^{+\infty} e^{i \nu v} \varphir_V(\gamma(v))dv,\\
&G_V(t,v,\nu,z) = e^{A(\tau,\nu,z) + vB(\tau,\nu,z)}, \label{eq:G_hat} \\
&A(\tau,\nu,z) = -\frac{2\kappa m}{\overline{\eta}^2(z)} \log\left(\nu\frac{\overline{\eta}^2(z)}{2\kappa}(1-e^{-\kappa \tau}) + 1 \right),\\
&B(\tau,\nu,z) = \frac{\nu e^{-\kappa \tau}}{\nu\frac{\overline{\eta}^2(z)}{2\kappa}(1- e^{-\kappa \tau}) + 1}.
\end{align*}
Furthermore, the Fourier transform of some typical payoff functions are
\begin{align}
\varphir_V(\gamma(v)) = \sqrt{m(1 - \theta) + \theta v} &\Rightarrow \widehat{\varphir_V}(\nu) = 2\pi^{3/2}\frac{ e^{-\frac{m(1-\theta)}{\theta}\nu}}{ \theta\left(-\dfrac{\nu}{\theta} \right)^{3/2}}, \\
\varphir_V(\gamma(v)) = (\sqrt{m(1 - \theta) + \theta v} -K)^+ &\Rightarrow \widehat{\varphir_V}(\nu) = \sqrt{\pi}\frac{ \left(1 - \erf\left(K\sqrt{-\dfrac{\nu}{\theta}} \right)\right)e^{-\frac{m(1-\theta)}{\theta}\nu}}{2 \theta\left(-\dfrac{\nu}{\theta} \right)^{3/2}},
\end{align}
where $\erf(z)$ is the complex error function, see \cite{sepp_vix_fourier}. For the VIX future case ($\varphir_V(v) = v$), one needs to remove the discount factor $e^{-r(T-t)}$ in $P_{V_0}$.

By equation (\ref{eq:P0_vix_formula}),
\begin{align}
&\frac{\partial^2 P_{V_0}}{\partial v \partial \eta}(t,v,z) = \frac{e^{-r\tau}}{\pi} \int_0^{+\infty} \left(\frac{\partial B}{\partial \eta} + B\frac{\partial A}{\partial \eta} + vB\frac{\partial B}{\partial \eta} \right) G_V(t,v,\nu, z)  \widehat{\varphir_V}(\nu) d\nu_i, \\
&\frac{\partial^3 P_{V_0}}{\partial v^3}(t,v,z) = \frac{e^{-r\tau}}{\pi} \int_0^{+\infty} B^3(\tau,\nu) G_V(t,v,\nu, z)  \widehat{\varphir_V}(\nu) d\nu_i.
\end{align}
We then consider the following educated guess:
$$P_{V_0} + P_{V_{1,0}}^{\eps} + P_{V_{0,1}}^{\delta} = \frac{e^{-r\tau}}{\pi} \int_0^{+\infty} Re\left((1 + h_{V_0}^{\eps,\delta} + v h_{V_1}^{\eps,\delta} + v^2 h_{V_2}^{\eps,\delta}) G_V(t,v,\nu)  \widehat{\varphir_V}(\nu) \right) d\nu_i,$$
where $h^{\eps,\delta}_{V_i}$ are functions of $\tau$, $\nu$ and $z$. By Equations (\ref{eq:P10eps_vix}) and (\ref{eq:P01delta_vix}), one may conclude that $h^{\eps,\delta}_{V_i}$ must solve the ODE system
\begin{align}\label{eq:ode_system_P_V}
\left\{\begin{array}{l}
\ds \frac{\partial h_{V_2}^{\eps,\delta}}{\partial \tau}(\tau,\nu,z) = 2(-\kappa + B(\tau,\nu) \overline{\eta}^2(z))h_{V_2}^{\eps,\delta}(\tau,\nu,z) + V_1^{\delta}(z) B(\tau,\nu) \frac{\partial B}{\partial \eta}(\tau,\nu),  \\ \\
\ds \frac{\partial h_{V_1}^{\eps,\delta}}{\partial \tau}(\tau,\nu,z) = (-\kappa + B(\tau,\nu) \overline{\eta}^2(z))h_{V_1}^{\eps,\delta}(\tau,\nu,z) + (2\kappa m + \overline{\eta}^2(z))h_{V_2}^{\eps,\delta}(\tau,\nu,z)\\ \\
 \ds \hskip 1cm + V_3^{\eps}(z) B^3(\tau,\nu) + V_1^{\delta}(z) \left(\frac{\partial B}{\partial \eta}(\tau,\nu) + B(\tau,\nu)\frac{\partial A}{\partial \eta}(\tau,\nu)\right), \\ \\
\ds \frac{\partial h_{V_0}^{\eps,\delta}}{\partial \tau}(\tau,\nu,z) = \kappa m h_{V_1}^{\eps,\delta}(\tau,\nu,z), \\ \\
h_{V_0}^{\eps,\delta}(0,\nu,z) = h_{V_1}^{\eps,\delta}(0,\nu,z) = h_{V_2}^{\eps,\delta}(0,\nu,z) = 0.
\end{array}
\right.
\end{align}

\section*{Funding}

J.-P. Fouque was supported by NSF grant DMS-1409434.

\bibliographystyle{plainnat}

\end{document}